\documentclass[fleqn,usenatbib]{mnras}
\hypersetup{draft} 

\usepackage{newtxtext,newtxmath}

\usepackage[T1]{fontenc}
\usepackage{ae,aecompl} 
\usepackage{lscape}

\usepackage{graphicx}	
\usepackage{amsmath}	

\newcommand{\kms}{km\,s$^{-1}$}

\title[Long-term stellar activity variations and their effect on radial-velocity measurements]{Long-term stellar activity variations and their effect on radial-velocity measurements}

\author[J. C. Costes et al.]{
\parbox{\textwidth}{
Jean C. Costes$^{1}$\thanks{E-mail: \href{jcostes01@qub.ac.uk}{jcostes01@qub.ac.uk}},
Christopher~A.~Watson$^{1}$,
Ernst~de~Mooij$^{1}$,
Steven H. Saar$^{2}$,
Xavier~Dumusque$^{3}$,
Andrew~Collier~Cameron$^{4}$,
David~F.~Phillips$^{2}$,
Maximilian~N.~G{\"u}nther$^{5}$\thanks{Juan Carlos Torres Fellow},
James~S.~Jenkins$^{6, 7}$,
Annelies~Mortier$^{8,9}$,
Andrew~P.~G.~Thompson$^{1}$
}
\vspace{0.3cm}
\\
$^{1}$Astrophysics Research Centre, School of Mathematics and Physics, Queen's University Belfast, BT7 1NN, Belfast, UK\\
$^{2}$Center for Astrophysics | Harvard \& Smithsonian, 60 Garden Street, Cambridge, MA 02138 USA\\
$^{3}$Observatoire Astronomique de l'Universit\'{e} de Gen\'{e}ve, 51 Chemin des Maillettes, 1290 Sauverny, Suisse\\
$^{4}$Centre for Exoplanet Science, SUPA School of Physics and Astronomy, University of St Andrews, North Haugh, St Andrews KY16 9SS, UK\\
$^{5}$Department of Physics, and Kavli Institute for Astrophysics and Space Research, Massachusetts Institute of Technology, Cambridge, MA 02139, USA\\
$^{6}$Departamento de Astronomia, Universidad de Chile, Casilla 36-D, Santiago, Chile\\
$^{7}$ Centro de Astrof\'isica y Tecnolog\'ias Afines (CATA), Casilla 36-D, Santiago, Chile.\\
$^{8}$Astrophysics Group, Cavendish Laboratory, University of Cambridge, J.J. Thomson Avenue, Cambridge CB3 0HE, UK\\
$^{9}$Kavli Institute for Cosmology, University of Cambridge, Madingley Road, Cambridge CB3 0HA, UK\\
}
\date{Accepted XXX. Received YYY; in original form ZZZ}

\pubyear{2020}

\begin{document}
\label{firstpage}
\pagerange{\pageref{firstpage}--\pageref{lastpage}}
\maketitle

\begin{abstract}
Long-term stellar activity variations can affect the detectability of long-period and Earth-analogue extrasolar planets. We have, for 54 stars, analysed the long-term trend of five activity indicators: log\,$R'_\mathrm{{HK}}$, the cross-correlation function (CCF) bisector span, CCF full-width-at-half-maximum, CCF contrast, and the area of the Gaussian fit to the CCF; and studied their correlation with the RVs. The sign of the correlations appears to vary as a function of stellar spectral type, and the transition in sign signals a noteworthy change in the stellar activity properties where earlier type stars appear more plage dominated. These transitions become more clearly defined when considered as a function of the convective zone depth. Therefore, it is the convective zone depth (which can be altered by stellar metallicity) that appears to be the underlying fundamental parameter driving the observed activity correlations. In addition, for most of the stars, we find that the RVs become increasingly red-shifted as activity levels increase, which can be explained by the increase in the suppression of convective blue-shift. However, we also find a minority of stars where the RVs become increasingly blue-shifted as activity levels increase. Finally, using the correlation found between activity indicators and RVs, we removed RV signals generated by long-term changes in stellar activity. We find that performing simple cleaning of such long-term signals enables improved planet detection at longer orbital periods.

\end{abstract}



\section{Introduction}
\label{sec:intro}

The latest highly wavelength-stabilised spectrographs (such as ESPRESSO, see e.g. \citealt{Gonzalez2018, Pepe2020}) can achieve hitherto unprecedented radial velocity (RV) instrumental precision. Within the context of exoplanet research, this should enable greater RV sensitivity to orbiting exoplanets through the study of the Doppler wobble reflex motion they induce on the host star. However, as such RV measurements become more precise they also become progressively more sensitive to the RV signals driven by stellar activity (see e.g. \citealt{Lovis2011, Dumusque2011a, Meunier2013}). This is a particular issue for the confirmation of long-period low-mass planets, which is becoming an increasing focal point of research as the field strives towards the discovery and confirmation of Earth-analogue planets.

Low-mass main-sequence stars, such as the Sun, generate magnetic activity that manifests itself as brighter or darker zones on their surface, such as spots or plage regions. Due to this stellar activity, the light emitted is affected and causes variations in the shape of the observed stellar spectral lines. For example, the contrast effect of starspots leads to an apparent emission `bump' in the stellar absorption line-profiles. While these bumps are typically not resolved directly in the absorption line profiles themselves, they manifest themselves as a change in the line asymmetries and induce RV shifts between a few $m~s^{-1}$ up to 100~$m~s^{-1}$ for the most active stars (e.g. \citealt{Saar1997, Hatzes2002, Desort2007}). The presence of faculae/plage, bright regions associated with elevated magnetic field, also induces a variable signal in the RVs. Since convection consists of bright, hot, up-flowing granules surrounded by dark, sinking, intergranular lanes, then the dominant spectral imprint of convection is a net blue-shifted component to the spectral line-profile. However, in the presence of plage, this convective blue-shift term is suppressed. This mechanism, considered to dominate the stellar induced RVs for low-activity stars \citep{Meunier2010}, is thought to be the biggest impediment to the RV detection of Earth-like planets. There are a number of other sources of apparent RV variations, including stellar oscillations (with RV variations between 0.10 to 4~$m~s^{-1}$ – e.g. \citealt{Schrijver2000, Chaplin2019}); meridional flows (between 0.7 to 1.6~$m~s^{-1}$, see \citealt{Meunier2020}); granulation (from a few $cm~s^{-1}$ up to $m~s^{-1}$ – e.g. \citealt{Schrijver2000, Dumusque2011, Meunier2015, Cegla2019, Cameron2019}); and supergranulation (between 0.3 to 1.1~$m~s^{-1}$, see \citealt{Dumusque2011, Meunier2015}).

If not properly accounted for, these line-profile shape changes may be mis-interpreted as shifts in the observed line's central wavelengths - which may then be mistaken for apparent radial-velocity shifts (see e.g. \citealt{Queloz2001, Desidera2004, Rajpaul2016}). Even for low-activity stars, this can have a dramatic impact on the radial-velocity follow-up used for determining the (minimum) mass of exoplanets. Given the long orbital periods of such planets, it is also possible that any RV study of such a system would need to span several years in order to robustly retrieve a repeating planetary signal. We therefore need to understand the impact of stellar activity over a range of different timescales. On timescales of days/weeks, stellar activity RV variations are predominantly driven by rotational modulation of spots and plage regions as they cross the visible stellar hemisphere. While a number of techniques and observational strategies have been developed to mitigate their effects on short timescales \citep{Dumusque2018}, monitoring stars on longer timescales (i.e. a substantial fraction of a stellar activity cycle) presents a different set of problems for stellar-activity mitigation techniques (see e.g. \citealt{Santos2010, Lovis2011}). Indeed, over long-timescales the activity-cycle will drive changes in the magnetic network and this will, in turn, change both the plage coverage \citep{Meunier2018, Meunier2019b} as well as the underlying magnetic network variation, and may affect surface flows \citep{Meunier2020}. In addition, stars may vary between exhibiting RV signals that are spot-dominated and those that are plage-dominated as the magnetic activity cycle waxes and wanes.

There are a number of astrophysical noise removal methods such as using wavelength dependent indices (e.g. \citealt{Anglada2012, Tuomi2013, Feng2017}), or Gaussian process regression (e.g. \citealt{Haywood2014, Rajpaul2015, Lopez2016}). This last one performs well in cases where the planetary orbital period is shorter than the typical active-region lifetime. However, it is likely that such technique will remove any longer period, low-amplitude planetary signals \citep{Langellier2020}. Other methods to remove long-term RV variations induced by stellar activity usually focus on a magnetic cycle correction of the RVs determined by fitting the log\,$R'_\mathrm{{HK}}$ variations with a polynomial, sinusoidal or Keplerian function \citep{Delisle2018, Diaz2018, Udry2019}. These correction methods, which are rarely used and only if the star presents an extremely large magnetic activity cycle variation, do not fully explore the stellar activity noise removal as they mostly concentrate on the trend with log\,$R'_\mathrm{{HK}}$, and are normally applied on a case-by-case basis.

The goal of this paper is to study long-term activity trends (as driven by, for example, stellar activity cycles) and their impact on measured RVs, with the ultimate goal of mitigating this impact to improve the detection of long-period exoplanets. We present a study of the long-term trends and their relationship to the measured RVs of 5 activity indicators: the log\,$R'_\mathrm{{HK}}$, the bisector span (BIS -- \citealt{Queloz2001}) of the cross-correlation function (CCF), the full width at half maximum (FWHM) of the CCF, the CCF contrast, and the area of the Gaussian used to fit the CCF. This study was done using observations of 54 stars, the Sun included, over long timescales (2 years for the Sun and up to $\sim$10 years for other stars) using data from HARPS and the HARPS-N solar telescope. The source of the data for our sample are described in Section~\ref{sec:observations}. Section~\ref{sec:indicators} outlines the different activity indicators used, the form of their long-term variations, and how they depend on one another as a function of stellar spectral type or as a function of the convective zone depth. In Section~\ref{sec:RVs}, we present how these activity indicators can influence the measured RVs by looking at the changes seen in the correlation between activity indicators and RVs and discuss their physical meaning. In Section~\ref{sec:Discussion}, we discuss a method to remove long-term RV trends due to the stellar activity cycle using the best activity indicators, with the aim of improving the detection capabilities of exoplanets in long-period orbits. Finally, we finish with our conclusions in Section~\ref{sec:conclusions}.



\section{Observations and data reduction}
\label{sec:observations}

We used publicly available data from the HARPS spectrograph \citep{Mayor2003} mounted on the ESO 3.6 m telescope at La Silla Observatory, Chile, to study the long-term variability of 53 apparently slowly-rotating stars (as measured by their log\,$R'_\mathrm{{HK}}$), hence we are selecting stars that should be less active. Additionally, we initially looked at some M-type stars. However the M2 line-mask has been found to heavily distort the CCFs (see e.g. \citealt{Rainer2020}). Moreover, we also found that the values of the FWHM obtained for stars where a M2 line-mask was used were systematically (and significantly) lower than for the stars where other line-masks were employed. While some methods have tried to correct the distortions in the radial-velocity and FWHM measurements \citep{Suarez2015}, the offset found in the FWHM is still present. We therefore decided to focus only on the F-, G- and K-type stars. 

The HARPS spectrograph covers a wavelength between 3780 to 6910~\AA\ with a resolving power of R = 115,000. The 53 stars were chosen based on the best availability of frequent sampling over many years on the ESO archive, sufficient to see the changes in the magnetic cycle of the stars. Thus, a minimum of 4 years (1500 days) of baseline was required, for which at least 5 seasons of data were available, and for each season at least 4 nights of observations also had to be available. Another consideration in the selection process was to remove fast rotating stars in order to focus on the observations of quiet, low activity, stars. We therefore only selected stars with estimated rotation periods (from their mean log\,$R'_\mathrm{{HK}}$ values and the relationship of \cite{Noyes1984}) longer than 20 days. This selection, however, was not applied to F-type stars, as they undergo less magnetic braking over their lifetime relative to later type stars. Instead, we confirmed that the F-type stars chosen were displaying relatively low levels of activity as measured by log\,$R'_\mathrm{{HK}}$. The full sample is presented in Table~\ref{tab:1}. We used the reduced data available on the ESO archive\footnote{ESO archive: \url{http://archive.eso.org/wdb/wdb/adp/phase3_main/form}}, which are processed with the HARPS-Data Reduction Software (DRS) pipeline.

For comparison, we also studied solar data, obtained from the public data from \citealt{Cameron2019}. HARPS-N is an echelle spectrograph \citep{Cosentino2012} located at the Telescopio Nazionale Galileo (TNG), at the Roque de los Muchachos Observatory, Spain. It covers the wavelength range between 3830 to 6930~\AA\ with a resolving power of R = 115000. With the custom-built solar telescope operating since 2015, HARPS-N provided disk-integrated solar spectra with a cadence of 5 minutes \citep{Dumusque2015, Phillips2016}, enabling Sun-as-a-star observations. It can typically observe the Sun for around 6 hours each day with an RV precision of approximately 40~$cm~s^{-1}$ \citep{Cameron2019}.

\subsection{Sample data}
\label{sub:reduction}
Table~\ref{tab:1} presents all the stars used in this study including their spectral type (which spans from F6V to K9V) and their B--V colour index data, as taken from SIMBAD\footnote{SIMBAD: \url{http://simbad.u-strasbg.fr/simbad/}}. Table~\ref{tab:1} also shows the total number of spectra used, the number of individual nights the object was observed as well as the total time-span covering all of the observations. Furthermore, Table~\ref{tab:1} presents the median activity level, the log\,$R'_\mathrm{{HK}}$ (this is not included in the ESO archive data and thus had to be measured from their reduced spectra as described in Section~\ref{sec:indicators}) and the minimum and maximum log\,$R'_\mathrm{{HK}}$ levels observed for each star. Finally, the stellar metallicity, the stellar mass and the stellar effective temperature (taken from different sources from VizieR\footnote{VizieR: \url{https://vizier.u-strasbg.fr/viz-bin/VizieR}, presented below Table~\ref{tab:1}}) are also presented in Table~\ref{tab:1}. All of the stars used in this study are main-sequence stars.

\subsubsection{HARPS data}
The radial velocities of 53 stars were derived from the standard HARPS post-reduction analysis, which consists of cross-correlating each echelle order with either a G2, K5 or M2 binary mask containing thousands of spectral lines, with the mask chosen to best correspond to the spectral type of the observed star. The cross-correlation functions (CCFs) produced for each order are then combined into a mean-weighted CCF from which the radial velocities are obtained by simply fitting a Gaussian to the CCF profile and measuring its centroid. These data were accessed via the ESO archive. The archival data were then vetted in order to ensure proper quality control. For example, in January 2015 there was an upgrade to the fibres requiring the HARPS vacuum enclosure to be opened. This intervention had an impact on the point spread function, and resulted in an offset between various measurements taken before and after the upgrade. Thus, in order to avoid systematic offsets in the data that may impact our analysis, we decided to only use data taken prior to 2,457,150 BJD (i.e. before the intervention to facilitate the fibre-upgrade). The aforementioned imposed date constraints leave us with a potential base-line of $\sim$10 years of data, which is on the order of the expected duration of one stellar-activity cycle for solar-type stars \citep{Baliunas1995}.

In addition, we also employed further quality control cuts to the remaining data to remove outlier points and/or those affected by low signal-to-noise or potential saturation. If the signal to noise ratio (SNR) for a spectrum was three times lower than its median value across all spectra, then this spectrum was removed. Similarly, if there was a risk of over-exposure, with a SNR value over 500, the data was also removed. Despite these quality-control cuts, some apparent outliers still remained. In order to identify and remove the most extreme outliers, we conducted a 'pre-screening' cut using a 30-$\sigma$ median absolute deviation (MAD) clip. We were able to identify that the majority of the points removed by this pre-screening were a result of observing the wrong target, based on a clear mismatch in the spectra. Once these extreme outliers were filtered out, we performed an additional cut using a 7-$\sigma$ iterative $4^\mathrm{th}$ order polynomial MAD fit clipping (without the polynomial fit being adversely affected by clusters of extreme outliers). This process was repeated until subsequent iterations did not remove any additional points. We opted for a conservative 7-$\sigma$ in order to avoid removing anything that may be astrophysical in nature, while still removing data-points that were clearly found to be outliers due to systematics. After investigation, we note that most of the outliers identified in this second 7-$\sigma$ cut were due to calibration and/or data reduction errors or twilight/moon contamination. These two cuts were applied on the radial-velocity data, as well as on the CCF FWHM, CCF contrast, CCF BIS, and on the log\,$R'_\mathrm{{HK}}$ data (presented in Section~\ref{sec:indicators}). From an initial 63,502 data points taken for our sample between the date-boundaries defined earlier, 6,101 were defined as having too low or overexposed SNR and 2,467 were removed as potential outliers.

Finally, after correcting our data for outliers, we removed (from the RVs) any known signals. First, we corrected the RVs for the secular acceleration for each star. The secular acceleration induces a signal in the RV measurements for stars with high proper-motions, and is caused by the changing radial and transverse components of the stellar velocity vector as the observed star passes by the Sun \citep{Kamp1986, Choi2013}:

\begin{equation}
   \mathrm{SA} = \frac{0.0229\times \mu^2}{\pi} \ (m\ s^{-1}\ yr^{-1})
\end{equation} where $\pi$ is the parallax in arcseconds and $\mu$ is the total proper motion in arcseconds per year, obtained from the GAIA DR2 \citep{GAIA2018} archive database\footnote{GAIA archive: \url{https://gea.esac.esa.int/archive/}}.

In addition, we subsequently removed known planetary signals from the corrected RVs, as we are only interested in the RVs induced by the stellar activity of the host-star. We used planetary catalogues, including exoplanet.eu\footnote{exoplanet.eu: \url{http://exoplanet.eu/catalog/}} and the NASA exoplanet archive\footnote{NASA exoplanet archive: \url{https://exoplanetarchive.ipac.caltech.edu/}}, to obtain the properties of exoplanets in the system, and compared those with any signals seen in the data. Signals identified as due to known planets or planet candidates were removed. However, since our data spans a longer timespan than most datasets used for determining the planetary orbits, we used \texttt{allesfitter}\footnote{allesfitter: \url{https://github.com/MNGuenther/allesfitter}} to model the signals, using the data from the literature as a starting point, in order to get updated parameters for known planets. For more information on \texttt{allesfitter}, we refer the reader to \citealt{allesfitter-code, allesfitter-paper}. The updated planet parameters will be published separately in another upcoming paper (Costes et al., in prep.).

Finally, one should note that while most of the stars presented in this paper are thought to be single stars, some of them are part of a binary or triple star system. This is the case for $\alpha$ Centauri B and HD65277, which have orbital periods of approximately 80 and 965~years, respectively. It was therefore necessary to remove an additional long-term RV trend stemming from the binary orbital motion. Following \cite{Dumusque2012} for $\alpha$ Centauri B, a second order polynomial fit was used to remove the RV effects caused by the companion star. Similarly, a linear trend was applied to HD65277. However, we note that by using a polynomial to correct for the binary motion, we may also slightly affect the RV variations due to the long-term magnetic cycle. Therefore, some caution is required when considering the results for these two stars (see Section~\ref{sec:RVs}).

\subsubsection{HARPS-N data}
In this work, the solar RVs used are those as measured using the HARPS-N Data Reduction System \citep{Sosnowska2012} as standard. The data used were taken between November 2015 and October 2017, just before a 3~month gap in the observations due to damage to the fibre coupling the solar telescope to the HARPS-N calibration unit \citep{Cameron2019}. In order to study the RVs induced only by solar activity, the Doppler wobble effects of the solar-system planets were removed using the JPL HORIZONS on-line solar system\footnote{JLP Horizons: \url{https://ssd.jpl.nasa.gov/horizons.cgi}}, a computational web-interface service that provides access to solar system data. Effects caused by differential atmospheric extinction were also removed from the RVs and exposures affected by clouds (which can produce a pseudo-Rossiter-McLaughlin effect as the cloud obscures differing portions of the solar disc) were identified using the HARPS-N exposure meter and also removed, as done by \citealt{Cameron2019}. The residual RVs, after following this process, have an rms amplitude of 1.6~$m~s^{-1}$, which is similar to those observed on stars with similar activity levels \citep{Isaacson2010}.


\section{Activity Indicators and long-term trends}
\label{sec:indicators}

In this Section, we present the different activity indicators used to study the long-term activity trend of the stars. As a first example, we look at the correlations between the log\,$R'_\mathrm{{HK}}$ and the CCF bisector span, FWHM, contrast, and the area for the Sun, and compare the results with the correlations found for $\alpha$ Centauri B. We then measure the correlations between activity indicators for our 54 stars, and present the different changes seen as a function of spectral type. These changes seem to unveil fundamental aspects for solar-type stars. Finally, studying these correlations as a function of convective zone depth (instead of the stellar effective temperature) provides further insight into the physical reasons behind the multiple transitions observed in the correlations between the different activity indicators.

\subsection{Measurements of log\,$R'_\mathrm{{HK}}$}
\label{sub:logRHK}

In the analysis presented in this paper, we have used the strength of the Ca\,II\,H\,\&\,K re-emission lines (as measured by log\,$R'_\mathrm{{HK}}$) as a tracer of stellar activity levels. Since the log\,$R'_\mathrm{{HK}}$ of each star in our sample is not a product of the ESO archive data we therefore measured these from the spectra, following the procedure outlined in \cite{Lovis2011}, which we summarise here. 

The S-index was determined by measuring the flux in two pass-bands centred on each of the Ca\,II\,H\,\&\,K line cores relative to two surrounding continuum regions (referred to as the V and R bands). The bands for the Ca\,II\,H\,\&\,K line cores are centered at rest wavelengths 3933.664\,\AA\ and 3968.470\,\AA, respectively, and have a triangular shape with a FWHM of 1.09\,\AA. The continuum V and R bands are centered at 3901.070\,\AA\ and 4001.070\,\AA, respectively, each with a width of 20\,\AA. The S-index, for the HARPS spectrograph, is then defined as:
\begin{equation}
    \mathrm{S}_{\mathrm{HARPS}} = \frac{H + K}{R + V}
\end{equation} where $H$, $K$, $R$ and $V$ represent the mean fluxes per wavelength interval for each band. This was measured for each star in our sample, and verified by comparing (where possible) with the results from \cite{Lovis2011}. We then used the calibration from \cite{Lovis2011} to map the S-index from HARPS to the Mt Wilson S-index:
\begin{equation}
    \mathrm{S}_{\mathrm{MW}} = 1.111\times \mathrm{S}_{\mathrm{HARPS}} + 0.0153
\end{equation} Then, after using a conversion factor, $C_\mathrm{cf}(B - V)$, that corrects for the varying flux in the continuum passbands, and after correcting for the photospheric contribution in the H\,\&\,K passbands, $R_\mathrm{phot}(B - V)$, we calculated the value of $R'_\mathrm{HK}$ defined by \cite{Noyes1984}, from the S-index:
\begin{equation}
\label{rhk_new}
    R'_{\mathrm{HK}} = 1.340\times 10^{-4}\times C_{\mathrm{cf}} \left(B - V \right)\times S_{\mathrm{MW}} - R_{\mathrm{phot}} \left(B - V \right)
\end{equation}

Finally, we compared the median log\,$R'_\mathrm{{HK}}$ of individual targets in our sample to those from \cite{Lovis2011} over the same timescale (where this was possible), finding a median error of less than $0.01$. This small discrepancy can be explained by the removal of different outliers in our work. Even though some small changes may affect the absolute values obtained, we are interested in the relative long-term variations in the log\,$R'_\mathrm{{HK}}$, which should still largely be preserved. In addition, while the same method was used to measure the log\,$R'_\mathrm{{HK}}$ of the Sun using HARPS-N data, we also removed the effects of spectrograph ghosts that are superimposed on the core of the Ca\,II\,H\,\&\,K lines to produce a high sensitivity and accuracy $R'_\mathrm{HK}$ \citep{Dumusque2020}.

\subsection{Other activity indicators and their long-term trends}
\label{sub:nwindicators}

Stellar activity, manifestations of which include the appearance of plage and/or spots, can change the shape of the spectral line as active regions evolve and/or rotate across the visible stellar disc. These line-shape changes are reflected in the CCF measurements as variations in the CCF bisector span, the CCF full-width-at-half-maximum (FWHM), and the CCF contrast. These CCF measures are products available in the ESO archive (or from the HARPS-N DRS in the case of the solar data) for our sample, and are generated using a spectral-type line-mask closely matched to the target's spectral-type (as explained in Section~\ref{sec:observations}).

It should be noted that both the CCF FWHM and CCF contrast (for the HARPS data) exhibit a slow, systematic long-term drift. This drift has been attributed to a change in the instrument focus with time \citep{Dumusque2018}. By comparing the drift for every star in our sample, we also noted an apparent colour-term dependency that impacts the shape and the amplitude of this instrumental drift. This colour-term could potentially arise in a number of ways. It could result from a variation in the instrumental focus across the chip, which would then manifest itself as a wavelength dependent change in the FWHM. This would be compounded by the varying spectral energy distribution of the targets and the different line masks used in the subsequent generation of the CCF. However, we believe that this effect may be primarily driven by differences in the projected rotation rates of the target stars. Stars with larger FWHMs are less affected by relatively small instrumental drifts, and since the earlier-type stars have higher projected rotational velocities, this could lead to an apparent colour-dependency in the long-term FWHM and contrast drifts.

Disentangling this drift from long-term stellar-activity variations of the FWHM and contrast is not straight-forward. While the observed colour-dependency is not strong, we still opted to apply different instrumental drift corrections by grouping stars that present broadly similar colour terms and intrinsic FWHMs. After investigation, we opted to correct the drift separately for each of the three broad spectral classes (F, G and K) in our sample. For each spectral class, we combined separately both the CCF FWHMs and contrast (after removing their mean value) of the stars and we simply fitted a 2$^\mathrm{nd}$ order polynomial to both the CCF FWHMs and contrast. By combining a large number of stars in each spectral class, it is hoped that any stellar activity-cycle variations largely average out, and that any underlying trend therefore follows the instrumental (and colour/intrinsic FWHM dependent) drift. We have then applied to the CCF FWHM ($\mathrm{FWHM'}$) and CCF contrast ($\mathrm{contrast'}$) the following drift corrections:
\begin{equation}
\begin{split}
    &\mathrm{FWHM} =~ \\
    &\mathrm{FWHM'}\,-\, \left(a\times \left(\mathrm{BJD} - \mathrm{BJD_{0}} \right)^2 + b\times \left(\mathrm{BJD} - \mathrm{BJD_{0}} \right) + c \right)
\end{split}
\end{equation}
\begin{equation}
\begin{split}
    &\mathrm{contrast} =~ \\
    & \mathrm{contrast'}\,-\, \left(d\times \left(\mathrm{BJD} - \mathrm{BJD_{0}} \right)^2 + e\times \left(\mathrm{BJD} - \mathrm{BJD_{0}} \right) + f \right)
\end{split}
\end{equation}

\begin{table*}
    \centering
    \begin{tabular}{c|c|c|c|c}
    \hline
    \hline
    	Correction  &   Coefficient &   F-type stars  &   G-type stars  &   K-type stars  \\
    \hline
    &   a   &   $8.3 \pm 1.3 \times 10^{-10}$   &   $9.4 \pm 0.7 \times 10^{-10}$   & $1.5 \pm 0.2 \times 10^{-9}$   \\
    FWHM    &   b   &  $1.0 \pm 0.7 \times 10^{-6}$   &  $2.6 \pm 0.3 \times 10^{-6}$   & $4.9 \pm 8.9 \times 10^{-7}$   \\
    &   c   &   $-1.01 \pm 0.08 \times 10^{-2}$   &   $-1.33 \pm 0.04 \times 10^{-2}$   & $-1.2 \pm 0.1 \times 10^{-2}$   \\
    \hline
    &    d   &  $7.3 \pm 0.8 \times 10^{-9}$   &  $7.0 \pm 0.6 \times 10^{-9}$   &  $-3.9 \pm 1.5 \times 10^{-9}$   \\
    contrast    &   e   &   $-4.9 \pm 0.4 \times 10^{-5}$ &   $-7.2 \pm 0.3 \times 10^{-5}$   & $-3.8 \pm 0.7 \times 10^{-5}$  \\
    &    f   &  $7.2 \pm 0.5 \times 10^{-2}$   &  $1.29 \pm 0.03 \times 10^{-1}$   &  $1.30 \pm 0.08 \times 10^{-1}$   \\
    \hline
      &   BJD$_{0}$   &   2452942.66   &   2452937.55   &   2452937.57  \\
    \hline
    \hline
    \end{tabular}
\caption{Values used for the correction of the instrumental (colour-term dependent) drift of the CCF FWHMs and contrasts. A different correction was made for each spectral class.}
\label{tab:2}
\end{table*}

 \begin{figure*}
    \centering
    \includegraphics[width=1\textwidth]{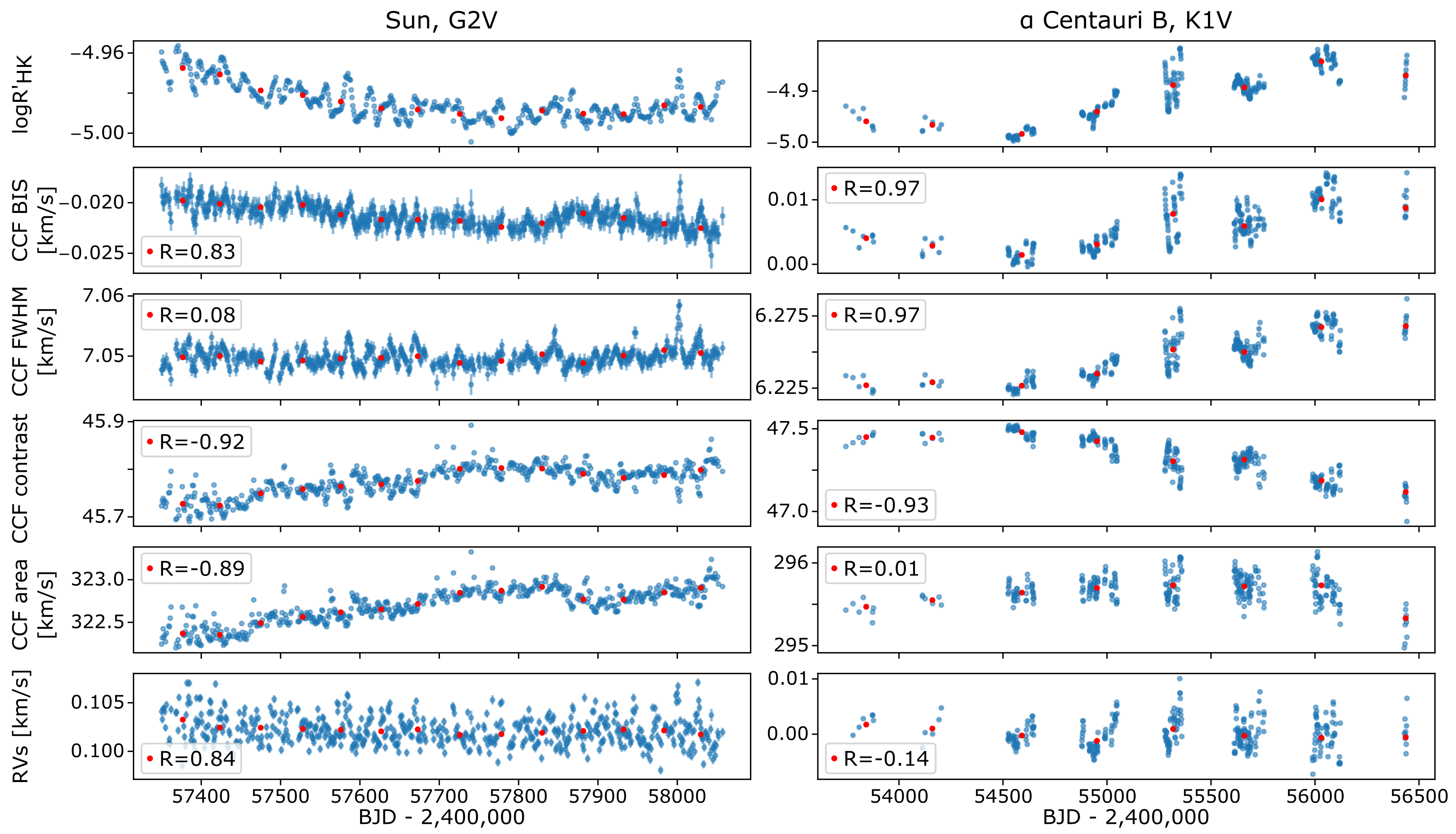}
    \caption{\label{fig:Correlation_1} The activity indicators and measured RVs for the Sun ({\it left}) and $\alpha$ Centauri B ({\it right}), spanning over 2 years and 10 years of HARPS-N and HARPS observations, respectively. The blue points correspond to the daily/nightly weighted mean, while the red points correspond to a 50 days weighted mean binning for the Sun and a seasonal weighted mean binning for $\alpha$ Centauri B. The Pearson's $R$ correlation coefficient, measured for each data set relative to log\,$R'_\mathrm{{HK}}$ for the binned data, is displayed in the legend. A net difference in the correlation between these two targets can be observed for the CCF FWHM and for the CCF area.}
\end{figure*}

\noindent The values of each coefficient for each spectral-type are presented in Table~\ref{tab:2}. We note that, while the drift we measured varies as a function of spectral type, the general form of the trend is not overly dissimilar across all types of stars in our sample -- that is, there is no sudden change in the shape of the trend as one looks at different stellar types/colours. Finally, following \cite{Cameron2019}, we have defined the CCF area as the product of the (corrected) CCF FWHM and contrast.

\subsection{Examples: The Sun and $\alpha$ Centauri B}
\label{subsub:SunandAlf}
In Figure~\ref{fig:Correlation_1} we present an example case study showing the 5 activity indicators and the RVs for the Sun and for $\alpha$ Centauri B, which cover a baseline of 2 and 10~years, respectively. These stars were chosen as they are extremely well-sampled and the progression through the stellar-activity cycle for $\alpha$ Centauri B is clear. The blue points correspond to the binned daily/nightly inverse-variance weighted mean of the measurements. Since we are interested in long-term activity changes, we have filtered out shorter period rotational modulation variations by binning each season of observations (again using a weighted mean). These are shown as red points for $\alpha$ Centauri B, and the same process was also followed for the other 53 HARPS targets in this paper. In the case of the solar data (which is continuously observed and therefore lacks clearly defined seasons) the binning was done every 50 days, corresponding to $\sim$2 solar rotations. As can be seen in Figure~\ref{fig:Correlation_1}, the red points nicely trace the long-term activity changes for both targets.

One can see the diminution of activity in the Sun, observed here as a decrease in log\,$R'_\mathrm{{HK}}$ in Figure~\ref{fig:Correlation_1}. This decline in activity seems also mirrored by a corresponding decrease in the BIS, thus, both of these activity indicators seem to be correlated over the long-term trend. While the FWHM looks flat, and does not seem to be strongly (anti-)correlated with long-term activity variations, both the contrast and the area are increasing as the solar activity level decreases. It is then safe to assume that, for the case of the Sun, it is the contrast that drives the long-term trend of the CCF area, compared to the FWHM, which has a smaller overall impact. In addition, we note a break in the trend lines of the BIS, contrast and area, around BJD 2,457,850. This interruption, observed by \cite{Cameron2019}, coincides with the appearance of large, persistent bipolar active regions in late March 2017. The trends are then resumed later from their new levels, around BJD 2,457,900. From the observations of isolated peaks in the FWHM, correlated with the number of spots, and the persistence of the BIS signal for several rotations, \cite{Cameron2019} suggested that the FWHM responds to dark spots while the BIS traces inhibition of convection in active-region faculae.

Finally, we can also see that the contrast is affected by both spots and plage regions. This is evidenced by the fact that we can see dips in the CCF contrast that correspond to sharp rises in the FWHM as spot regions rotate into view. The fact that, in addition, the CCF contrast shows appreciable long-term variability indicates that it is also impacted by the under-lying magnetic network. As a result of the interplay between the contrast and FWHM, the CCF area displays little evidence of rotational modulation as the variation of the FWHM driven by spots appears to be largely cancelled out by the variation in the contrast. However, on longer timescales, the area shows an anti-correlation with log\,$R'_\mathrm{{HK}}$ indicating that it is long-term variations in the contrast that acts as the primary driver of changes in the CCF area in the case of the Sun.

When comparing the activity indicators trends of the Sun with those of $\alpha$ Centauri B, some significant differences are observed. In order to provide a statistical measure of the correlation strength for each indicator, the Pearson's $R$ correlation coefficient between the long-term variations of the CCF bisector span, FWHM, contrast, and the area (red points in Figure~\ref{fig:Correlation_1}) and log\,$R'_\mathrm{{HK}}$ is also shown in the inset panels on Figure~\ref{fig:Correlation_1} as well. Similar to the Sun, $\alpha$ Centauri B presents a strong correlation between log\,$R'_\mathrm{{HK}}$ and the BIS, and a strong anti-correlation between log\,$R'_\mathrm{{HK}}$ and the contrast. Concerning the FWHM, while it appears mostly flat for the Sun, it is strongly correlated in the case of $\alpha$ Centauri B. Moreover, instead of an anti-correlation between log\,$R'_\mathrm{{HK}}$ and the area, here we see a positive (albeit weak) correlation. This would suggest that in the case of $\alpha$ Centauri B, it is the FWHM that slightly drives the long-term trend of the area, instead of the contrast. Here we have only presented $\alpha$ Centauri B and the Sun as together they provide two clear examples where different long-term relationships between key activity indicators are readily identifiable (presumably as a result of their different spectral-types). In the rest of this paper, we present a more general investigation of how the stellar activity indicators vary as a function of fundamental stellar parameters (such as spectral type) using a sample of 54 stars.

 \begin{figure*}
    \centering
    \includegraphics[width=1\textwidth]{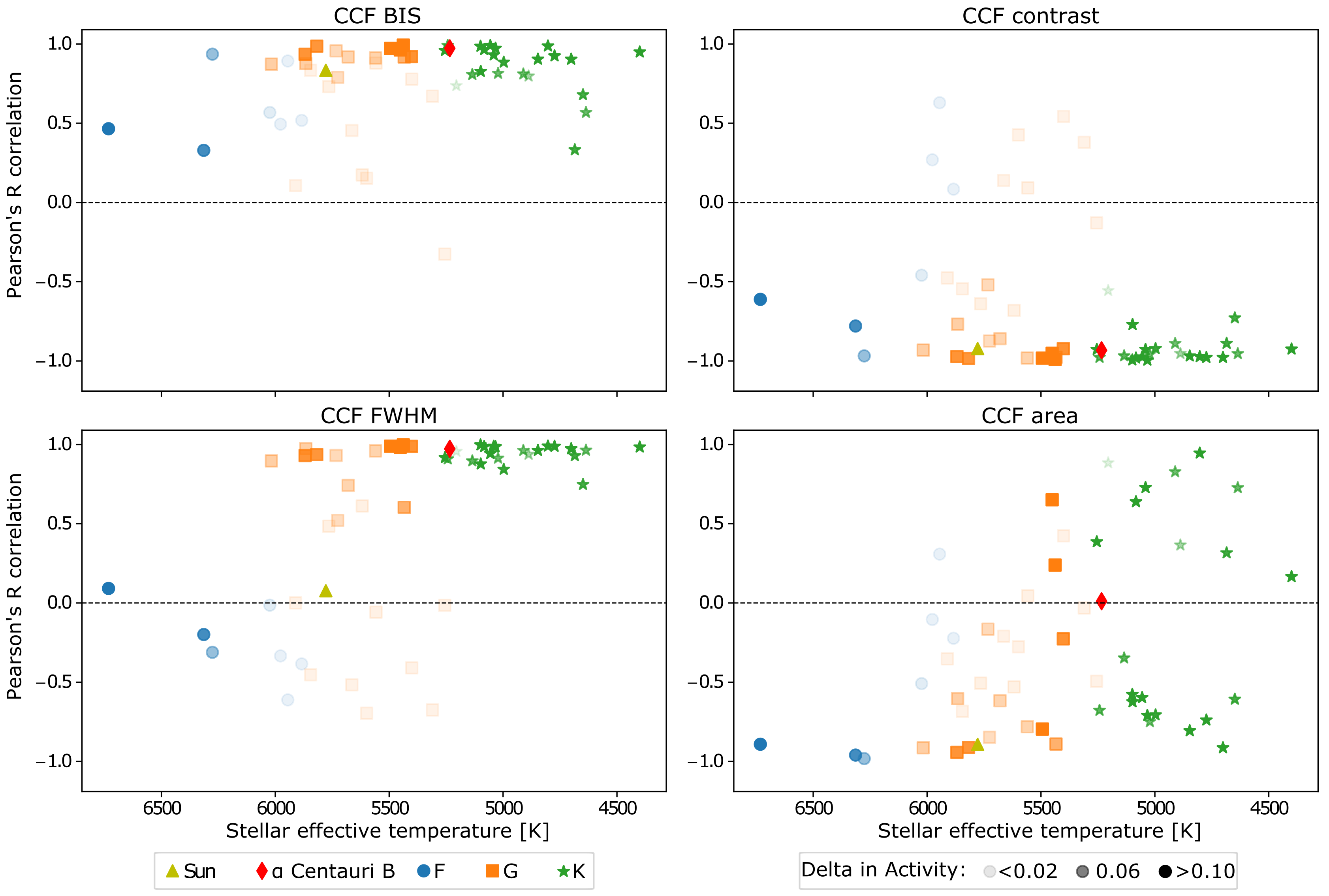}
    \caption{\label{fig:Correlation_2} Pearson's $R$ correlation between the log\,$R'_\mathrm{{HK}}$ and four activity indicators: CCF BIS (top left), CCF contrast (top right), CCF FWHM (bottom left) and CCF area (bottom right), as a function of the stellar effective temperature for the 54 stars observed. The Pearson's $R$ correlation was measured using the seasonal averaged data points. A different color and shape was assigned for each spectral class. The Sun and $\alpha$ Centauri B are highlighted in yellow and red, respectively. Transparency was used to indicate the long-term change in stellar activity ($\Delta_{\mathrm{act}}$). This was used in order to identify possible case of bias. Hence, a more transparent point should be treated more carefully, as it does not exhibit a strong trend.}
\end{figure*}

\subsection{(Anti-)correlation between activity indicators as a function of spectral type}
\label{transition}

We have repeated the analysis presented in Section~\ref{subsub:SunandAlf} (for the Sun and $\alpha$ Centauri B) for the other stars in our sample. Figure~\ref{fig:Correlation_2} represents the value of the Pearson's $R$ correlation coefficient for the seasonal weighted-mean data between the log\,$R'_\mathrm{{HK}}$ and each of the four activity indicators (CCF BIS, FWHM, contrast and area) as a function of stellar effective temperature for our 54 targets. We have also plotted the correlations as a function of colour index (B-V), and find this makes no difference to the global picture described here. For clarity, we use a different colour and shape for each spectral class, and the Sun and $\alpha$ Centauri B (whose detailed plots are shown in Figure~\ref{fig:Correlation_1}) are highlighted in yellow and red, respectively. Finally, the change in activity ($\Delta_{\mathrm{act}}$), which represents the difference between the highest and lowest activity level of a star measured using the seasonally binned log\,$R'_\mathrm{{HK}}$ data, was also determined for each star. This $\Delta_{\mathrm{act}}$ was measured such that:
\begin{equation}
    \Delta_{\mathrm{act}} = \mathrm{max}\left(\overline{\mathrm{log}\,R'_\mathrm{{HK}}}\right) - \mathrm{min}\left(\overline{\mathrm{log}\,R'_\mathrm{{HK}}}\right)
\end{equation}
where $\overline{\mathrm{log}\,R'_\mathrm{{HK}}}$ corresponds to the log\,$R'_\mathrm{{HK}}$ of the seasonally binned data. This $\Delta_{\mathrm{act}}$ is represented in Figure~\ref{fig:Correlation_2} with transparency and is shown in order to identify possible cases of bias. Indeed, small values of $\Delta_{\mathrm{act}}$ indicates that the target does not display a prominent activity-cycle variation in the log\,$R'_\mathrm{{HK}}$ and thus, one should be careful when interpreting the results for these targets.

We would like to stress that, while Figure~\ref{fig:Correlation_2} presents the Pearson's $R$ correlation coefficient between log\,$R'_\mathrm{{HK}}$ and the other stellar activity indicators, our main focus is not the exact value of the correlation coefficient itself. Rather, we use the Pearson's $R$ correlation coefficient as a relatively crude tool to simply determine whether any correlation is present, and if so, whether it is strong or weak. The reason why we urge caution against over-interpreting the coefficients obtained is because the Pearson's $R$ assesses linear relationships between two quantities. In the case of correlations between two stellar activity indicators, there is no particular reason to expect a linear relationship between the two. Indeed, any relationship between two activity indicators could be rather complex and also change its form as a function of activity level. For example, an increase in activity may cause a star to transition from plage-dominated to spot-dominated, which may in turn cause a sharp transition in how different stellar activity indicators behave. However, we should also note that we have performed the same analysis using the Spearman correlation (which assesses monotonic relationships) and similar results were found. 

From inspection of Figure~\ref{fig:Correlation_2} it is clear that, for most of the stars, the log\,$R'_\mathrm{{HK}}$ and CCF BIS are strongly correlated with one another. We also clearly find an anti-correlation between the log\,$R'_\mathrm{{HK}}$ and the CCF contrast for most of the stars, while the few stars that do not exhibit such strong anti-correlation have a low $\Delta_{\mathrm{act}}$. This agrees with the results found for $\alpha$ Centauri B and the Sun, as presented in Figure~\ref{fig:Correlation_1}. However, the picture regarding the correlation between FWHM and log\,$R'_\mathrm{{HK}}$ is more complex, and a larger spread is evident for hotter stars. Most F- and some G-type stars show either a weak correlation or a strong anti-correlation between FWHM and log\,$R'_\mathrm{{HK}}$, while late G- and K-type stars are strongly correlated. Thus, a transition in the correlation seems to occur around G-type stars. This could explain why the Sun, which is in the middle of this apparent transition, exhibits a lower correlation compared to $\alpha$ Centauri B (for which the FWHM is strongly correlated with the activity). Finally, for the CCF area, most of the hotter stars (including the Sun) show an anti-correlation with log\,$R'_\mathrm{{HK}}$, while cooler stars exhibit both anti-correlations and correlations. This bifurcation in the correlations, which occurs for late G- and early K-type stars, tallies with what we described earlier for $\alpha$ Centauri B see Figure~\ref{fig:Correlation_1}), which lies at the edge of this bifurcation and presents a low correlation between log\,$R'_\mathrm{{HK}}$ and the CCF area.

To confirm that the changes seen in the correlation are astrophysically driven and not due to an instrumental or computational issue, we analysed our data in more detail. We focused our attention on the region where there is a bifurcation observed for the correlation between CCF area and log\,$R'_\mathrm{{HK}}$, which occurs around late G and early K spectral-types. This also corresponds to the region where the line-mask used for the CCF changes from a G2V mask to a K5V mask. Thus, to verify that the difference in the correlation properties observed is not related to a change in the line-mask used to form the CCFs, we tested the impact of using an `incorrect' mask for some stars (i.e. we used a K5 mask for G-type stars and a G2 mask for K-type stars). While we noticed small changes in the absolute values of the CCF parameters measured, the overall long-term trends for the CCF BIS, FWHM and contrast were very similar between masks, and their correlation with log\,$R'_\mathrm{{HK}}$ did not change significantly. This demonstrates that the choice of mask is not responsible for the trends seen for these parameters in Figure~\ref{fig:Correlation_2}.

However, we did notice a change in the observed CCF area (the product of the corrected FWHM and contrast) when using a different mask. For many K-type stars, its correlation with log\,$R'_\mathrm{{HK}}$ changes from an anti-correlation to a correlation when a G2 mask is used. For a few G-type stars that presented a positive correlation between the CCF area and the log\,$R'_\mathrm{{HK}}$, an anti-correlation is observed when a K5 mask is used. These differences can be seen in Figure~\ref{fig:Correlation_2WM}, which represents the Pearson's $R$ correlation between the log\,$R'_\mathrm{{HK}}$ and the four activity indicators when the incorrect line-mask is applied, and can be directly compared to Figure~\ref{fig:Correlation_2}. In summary, when using the wrong line-masks, the relationship between CCF area and log\,$R'_\mathrm{{HK}}$ becomes more strongly anti-correlated for the hotter G-type stars (but broadly in-keeping with what was seen before). However the bifurcation seen for the cooler stars becomes less distinct with the anti-correlated stars generally flipping to positive correlations.

The changes seen when using an `incorrect' mask can be explained by the differences in the masks themselves. Indeed, not only does the K5 mask have twice as many lines as the G2 mask, but the weights used for lines common to both masks are also different. As a result, for any given star, using a K5 mask results in lower CCF FWHMs and lower CCF contrasts relative to the results using a G2 mask. As an example, Figure~\ref{fig:Correlation_2GK} compares the CCF parameters (as well as the log\,$R'_\mathrm{{HK}}$ and the RVs) of a star when using a K5 and G2 mask. While the long-term trends of the CCF FWHM and contrast are not impacted by the use of an incorrect line-mask, their absolute values are. This, in turn, changes the effective weighting of the contributions of the FWHM and contrast to the calculation of the CCF area, and it is this that drives the change in the correlation between the CCF area and log\,$R'_\mathrm{{HK}}$.

In summary, using an incorrect line-mask has little to no impact on the long-term correlations between log\,$R'_\mathrm{{HK}}$ and CCF FWHM, contrast, and BIS. For the hotter stars in our sample, this story is also largely true for the CCF area versus log\,$R'_\mathrm{{HK}}$. The only change is that for cooler stars this becomes more strongly positively correlated. We note that while the CCF area/log\,$R'_\mathrm{{HK}}$ bifurcation changes into a transition when the `wrong' mask is used, it still signifies a change dominated by astrophysical processes.

Finally, we have also investigated other potential sources of bias, such as the number of spectra, number of nights and the SNR of observations for individual stars. None of these could explain the bifurcation in the correlation between log\,$R'_\mathrm{{HK}}$ and the CCF area around the late G and early K spectral type. Thus, we concluded that the changes in the correlation are real, and likely due to physical changes occurring within the stars, such as changes in the magnetic network. This is discussed in more detail in Section~\ref{PhysicalMeaning}.

\subsection{Physical interpretations of the observed correlations}
\label{PhysicalMeaning}

As seen in Figure~\ref{fig:Correlation_2}, the majority of the stars show a correlation between log\,$R'_\mathrm{{HK}}$ and the CCF BIS. This is expected, as an increase in activity will broaden the bisector span since the BIS is sensitive to velocity suppression and traces inhibition of convection in active regions \citep{Meunier2017, Cameron2019}. In the case of log\,$R'_\mathrm{{HK}}$ versus CCF contrast, the anti-correlation observed is also expected, as an increase in activity will increase the plage coverage -- thereby boosting the continuum and decreasing the contrast (see e.g. \citealt{Cegla2013}). In addition, both spots and plage create apparent `emission' bumps in the stellar line profiles (see, for example, \citealt{Thompson2017}), which also acts to decrease the contrast. While the correlation between the log\,$R'_\mathrm{{HK}}$ and both the CCF BIS and CCF contrast exhibit the same behaviour for all spectral types, changes in the correlation behaviour between the log\,$R'_\mathrm{{HK}}$ and the other two activity indicators are evident for different spectral types. We will now discuss these in the next subsections.

\subsubsection{The CCF FWHM versus log\,$R'_\mathrm{{HK}}$}
\label{sub:fwhm_rhk}
As observed in Figure~\ref{fig:Correlation_2}, the CCF FWHM presents low R-values with log\,$R'_\mathrm{{HK}}$ for many F- and early G-type stars, Sun included, and a strong positive correlation for later-type stars. While the FWHM can be affected by an imperfect drift correction, our data strongly suggest that the change in correlation seen reflects a physical difference in stellar properties. Some of these weakly correlated and all of the strongly anti-correlated stars have rather low $\Delta_\mathrm{act}$, and we think that this point can explain the transition seen in the correlation. Indeed, the shallow convective outer-envelopes of hotter stars results in a relatively weak magnetic flux that is insufficient to create significant starspots. Thus, these stars will mostly be plage-dominated. As for the FWHM, it is expected to increase with the apparition of spots. This is due to the reduction of the normalized line profile, and hence, the increase in the line-width to compensate and approximately preserve the area \citep{Vogt1983, Cameron2019}. Therefore, the FWHM can be considered as an indicator of spots for solar-type stars, as observed for the Sun \citep{Cameron2019}. It is then natural to see, for plage-dominated stars, such weak (anti-)correlation between the FWHM and log\,$R'_\mathrm{{HK}}$. These observations agree with SOAP 2.0 simulations \citep{Dumusque2014}, which can be used to distinguish between stars that are spots or plage regions dominated. The transition to a strong correlation, happening for later-type stars, thus indicates an increase of spots on the stellar surface. Since we have preferentially selected a sample of inactive stars that we assume are all relatively slowly rotating, this increase in activity level is likely driven by the increase in the convective zone depth (see Section~\ref{sec:convectivezonedepth}).

While the change in the magnetic cycle of the hotter stars is small, as represented by a low $\Delta_\mathrm{act}$, most of these F- and early G-type type stars still predominantly present an anti-correlation between the FWHM and log\,$R'_\mathrm{{HK}}$. This dominant trend can be explained by the increased temperature in plage, which is weakening typical metal lines, making them slightly de-saturated (narrower). This anti-correlation seen for hotter stars was also confirmed by \cite{Beeck2015}, who used magnetohydrodynamic simulations to show the effect of the magnetic field on stellar line profiles as a function of limb-angle for different spectral types. They investigated the differences in the FWHM between simulations with B$_{0}$ = 500~G and non-magnetic simulations for the magnetically sensitive Fe I line at 6173~\AA. While we acknowledge that the CCF FWHM presented in this work is derived from many 1,000's of stellar lines, and the model only studied the magnetic response of one line, it is still useful as a means to provide the basis for a physical interpretation of our results.

Indeed, the model shows similarity with our observations, where a difference is observed in the FWHM between F3V and G2V stars. According to \cite{Beeck2015}, this difference is explained by the fact that for hotter F-type stars the FWHM tends to decrease as magnetic activity increases, compared to cooler stars where the FWHM is strongly enhanced for the line studied. The anti-correlation between the FWHM and the activity is due to more significant line weakening in F stars from the ``hot-wall" effect arising in plage. \cite{Lovis2011} also found similar results, observing long-term correlations between the $R'_\mathrm{HK}$ and the FWHM. According to their observations, for a given variation in $R'_\mathrm{HK}$, while the contrast sensitivity is roughly constant, the FWHM is more sensitive to activity variations in cooler stars. This again agrees with the transition in the correlation between the FWHM and log\,$R'_\mathrm{{HK}}$ seen in Figure~\ref{fig:Correlation_2}.

\subsubsection{The CCF area versus log\,$R'_\mathrm{{HK}}$}
\label{sec:area_rhk}

Similar to what was observed between the log\,$R'_\mathrm{{HK}}$ and the FWHM, the CCF area also shows a change in the sign of the correlation as a function of stellar effective temperature. Figure~\ref{fig:Correlation_2} shows an anti-correlation between log\,$R'_\mathrm{{HK}}$ and the CCF area for hotter stars. Interestingly, for cooler stars such as late G- and K-type stars, a bifurcation in the correlation between log\,$R'_\mathrm{{HK}}$ and the CCF area is evident, with some stars showing a correlation between these two parameters, and others an anti-correlation.

In their models, \cite{Beeck2015} also looked at the difference in the equivalent width (EW), which is equivalent to our CCF area, for simulations with B$_{0}$ = 500~G and the respective non-magnetic one for the magnetically sensitive Fe I line at 6173~\AA. The models show a transition in the response of the EW to activity occurring between G2V and K5V stars. According to \cite{Beeck2015} this change arises due to the fact that the EW tends to decrease with increasing magnetic activity for the F3V and G2V simulations, while for cooler stars it is strongly enhanced. The anti-correlation seen for hotter stars is again due to line weakening in the bright magnetic flux concentrations. \cite{Beeck2015}'s results can, therefore, partially explain why the correlation behaviour changes as a function of spectral-type (as we also see in Figure~\ref{fig:Correlation_2}), where some cooler stars present a positive correlation between the CCF area and the log\,$R'_\mathrm{{HK}}$. However, other cooler stars do not follow this trend and, instead exhibit an anti-correlation. 

In order to further investigate the underlying cause of this bifurcation seen for K-type stars, we analysed whether the inclination of the stars could have an impact on the observed correlations. Inclination could, potentially, explain some of the bifurcation behaviour seen in Figure~\ref{fig:Correlation_2} as a result of differing line-of-sights to active regions. Unfortunately, it is not possible to measure robustly the inclinations of the stars in our sample. Instead, since we drew our sample from a population of assumed slow rotators, we have examined whether or not stars lying on the upper or lower arm of the bifurcation show stronger rotationally modulated signals in periodogram analyses of their activity indicators. Since stars viewed edge-on should show larger rotational modulation of stellar features (as active regions rotate on-and-off the visible stellar disc) compared to stars viewed more pole-on, we wished to see whether stars on one arm preferentially showed stronger periodogram peaks around plausible stellar rotation periods (or associated harmonics). However, no such difference was found in the general structure of the periodograms of stars on either bifurcation arm, and therefore we cannot establish any evidence that this bifurcation is due to an inclination dependency.

We also investigated whether stars on either side of the bifurcation showed different general activity levels (as measured by their median log\,$R'_\mathrm{{HK}}$). On average, the K-type stars showing a positive correlation between the CCF area and log\,$R'_\mathrm{{HK}}$ were indeed more active, with a median log\,$R'_\mathrm{{HK}}$ = $-4.896 \pm 0.030$, compared to their anti-correlated counterparts (log\,$R'_\mathrm{{HK}}$ = $-4.993 \pm 0.019$). We therefore hypothesize that the bifurcation is due to differences in the activity levels between the two groups of stars, and suggest that the more active stars that show strong correlations between the CCF area and log\,$R'_\mathrm{{HK}}$ have a higher spot-coverage compared to those that show strong anti-correlations. However, a second hypothesis could also explain the bifurcation seen. If we consider the CCF area changing as a whole, then this implies that the stellar effective temperature is changing, most likely due to changes in the magnetic network coverage. In turn, this then implies that the CCF area is actually acting as a global activity-cycle indicator of the network coverage. In this case, the bifurcation could be due to a different magnetic network configuration or response over the activity cycle related to the depth of the convective zone. We think that it is a combination of both mechanisms (increase in spot-coverage and changes in the magnetic network) that impact the correlation between the CCF area and the log\,$R'_\mathrm{{HK}}$, leading to the bifurcation seen. In conclusion, for long timescales (longer than rotational modulations) and contrary to what is seen on the Sun, for the more active and cooler stars, it is long-term variations in the FWHM that acts as the primary driver of changes in the CCF area.

 \begin{figure}
    \centering
    \includegraphics[width=\columnwidth]{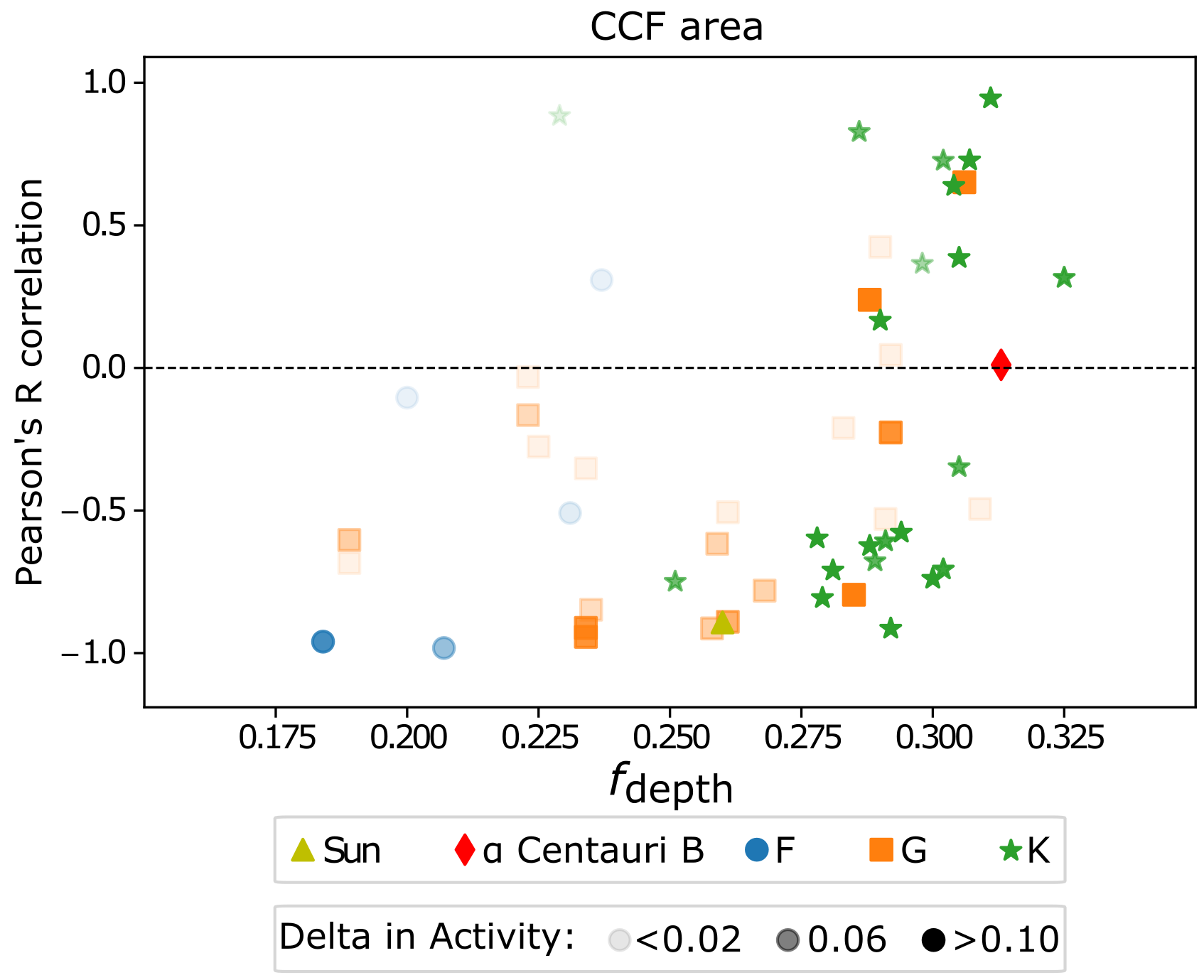}
    \caption{\label{fig:Correlation_22} Correlation between the CCF area and log\,$R'_\mathrm{{HK}}$ as a function of the fractional convective zone depth ($f_\mathrm{depth}$). Note that the x-axis is restricted to capture the bulk of the stars in our sample that exhibited the bifurcation behaviour discussed in Section~\ref{sec:area_rhk}. Therefore, some F-type stars, with more extreme fractional convective zone depth, were removed for this plot. Similar to Figure~\ref{fig:Correlation_2}, the Pearson's $R$ correlation was measured using the seasonal data points. Transparency effect, colours and shapes are the same as in Figure~\ref{fig:Correlation_2}.}
\end{figure}

\subsubsection{Potential metallicity and convective zone depth effects}
\label{sec:convectivezonedepth}

In Section~\ref{sub:fwhm_rhk}, we suggested that the smaller convective envelope of hotter stars results in a weaker magnetic field, unable to sustain substantial spot filling-factors on their surfaces. This, in turn, could explain the weak correlations and anti-correlations found between the CCF FWHM and log\,$R'_\mathrm{{HK}}$ for these early-type stars. This is in contrast to later spectral-types that largely exhibit a strong correlation between FWHM and log\,$R'_\mathrm{{HK}}$, which we hypothesize is a result of higher spot-filling factors. However, there are a few exceptions, where some cooler stars also exhibit a much weaker correlation than other stars with a similar $T_{\mathrm{eff}}$ (see Figure~\ref{fig:Correlation_2}). One of these coolest exceptions that exhibits an anti-correlation between FWHM and log\,$R'_\mathrm{{HK}}$ (rather than a correlation) is HD10700 ($\tau$ Cet), a G8V star with T$_{\mathrm{eff}}$ = 5310~K. Interestingly, this star also stands out for its particularly low metallicity of [Fe/H]~=~-0.52 \citep{Costa2020}.

Since metallicity affects the depth of the outer convective envelope of stars (see e.g. \citealt{Saders2012, Tanner2013}) and, in turn, the convective zone depth is a key ingredient of the magnetic-field generating stellar dynamo, we decided to investigate how activity indicators behave as a function of convective zone depth. Focusing on $\tau$ Cet as an example, we used EZ Web\footnote{EZ-Web: \url{http://www.astro.wisc.edu/~townsend/static.php?ref=ez-web}}, a web-based interface to the Evolve ZAMS (EZ) code \citep{Paxton2004}, to model its stellar structure. We define the fractional convective zone depth ($f_\mathrm{depth}$) as:
\begin{equation}
    f_\mathrm{depth} = \frac{R_\mathrm{s} - r}{R_\mathrm{s}}~,
\end{equation}
\noindent where $R_\mathrm{s}$ corresponds to the stellar radius given by the EZ-Web model, and $r$ is the depth of the convective zone as given by the model.

Keeping the stellar mass fixed, we calculated the fractional convective zone depth of $\tau$ Cet for a low metallicity ([Fe/H] = $-0.52$) and for a solar metallicity. For the low metallicity case we find that $f_\mathrm{depth}$ = ~0.223, whereas $f_\mathrm{depth}$ = 0.318 for the solar metallicity case. Hence, the extremely low metallicity of $\tau$ Cet seems to result in a much shallower convective zone depth than if it had solar metallicity. Looking at Figure~\ref{fig:Correlation_2}, this low metallicity has the effect of essentially shifting $\tau$ Cet further to the left in terms of its behaviour, more similar to that of an early G-type star with solar metallicity (for comparison, the fractional convective zone depth of the Sun is 0.26). We therefore believe that taking into account the metallicity can then bring the log\,$R'_\mathrm{{HK}}$/FWHM anti-correlation behaviour of $\tau$ Cet in-line with expectations.

From these observations, we then estimated the fractional convective zone depths of all the stars in our sample. For the purposes of this work, we have assumed ages of 4~Gyrs for the F-stars in our sample, and 5~Gyrs for other spectral-types. These ages equate to old (for F-stars) and/or solar-like ages, and we have assumed these as they will correspond to fairly inactive stars, broadly reflecting the sample of stars in this study. We note that we also examined the impact of varying the assumed ages by 1~Gyr, and we find that it does not impact on the findings presented forthwith. The stellar masses and metallicities used for the models (plus the source of this information) are presented in Table~\ref{tab:1}.

In Section~\ref{sec:area_rhk} we discussed a bifurcation in the correlation between CCF area and log\,$R'_\mathrm{{HK}}$ for later G- and K-type stars, where some displayed strong correlations, while others showed strong anti-correlations. We noted that, on average, those stars on the bifurcation arm that showed a strong correlation between CCF area and log\,$R'_\mathrm{{HK}}$ were more active than those stars that showed an anti-correlation.
Figure~\ref{fig:Correlation_22} presents the Pearson's $R$ correlation between the CCF area and the log\,$R'_\mathrm{{HK}}$, but now as a function of the fractional convective zone depth, using the same colour/transparency scheme as in Figure~\ref{fig:Correlation_2}. For clarity, we have restricted the range in convective zone depth in Figure~\ref{fig:Correlation_22} to focus on the main bulk of stars in our sample that exhibited the aforementioned `bifurcation'. By replacing the stellar effective temperature with the fractional convective zone depth, instead of the bifurcation observed previously, we now see a clear transition happening in the correlation around a fractional convective zone depth of $\sim$0.30 with $\alpha$ Centauri B (highlighted in red) lying around the center of this transition.

Therefore, we find that the K-type stars that present a positive correlation between the CCF area and the log\,$R'_\mathrm{{HK}}$ have deeper convective envelopes. In addition to being more active (with a median activity level of -4.896 ± 0.030, see Section~\ref{sec:area_rhk}), these stars are also, on average, more metal rich, with a median metallicity of $-0.01 \pm 0.07$. In comparison, the stars that show an anti-correlation between the CCF area and the log\,$R'_\mathrm{{HK}}$ are less active (with a median activity level of -4.993 ± 0.019), and have a lower metallicity (with a median metallicity of $-0.32 \pm 0.05$). This difference in metallicity and activity level between the stars lends more evidence to our previous hypothesis from Section~\ref{sec:area_rhk}. If the ages and masses are the same, then we observe that more metal rich stars have a deeper convective zone depth, which leads to a higher activity level. In turn, we speculate that this results in higher spot-coverage and hence the FWHM becomes the dominant factor driving the variation in the CCF area (over the contrast), resulting in a positive correlation between CCF area and log\,$R'_\mathrm{{HK}}$.

To conclude, Section~\ref{sec:indicators} unveils two fundamental aspects for solar-type stars.
\begin{enumerate}
\item Due to their shallow convective outer-envelopes, hotter F- and G-type stars present a low Pearson's $R$ correlation coefficient between the CCF FWHM and log\,$R'_\mathrm{{HK}}$. This thus suggests that these hotter stars are mostly plage-dominated. In comparison, the cooler stars present a strong correlation between the CCF FWHM and log\,$R'_\mathrm{{HK}}$, indicating the increase of spots on the stellar surface.
\item Cooler, late G- and K-type stars can be classified into two groups. The first group consists of more metal rich and more active stars with deeper convective envelopes. These stars can be identified by their positive correlation between the CCF area and the log\,$R'_\mathrm{{HK}}$, which suggests a higher spot-coverage. The second group consists of more metal poor and less active stars with shallower convective outer-envelopes. They can be identified by an anti-correlation between the CCF area and the log\,$R'_\mathrm{{HK}}$, which suggests that the stars are mostly plage-dominated.
\end{enumerate}


\section{The impact of Stellar Activity on Radial Velocities}
\label{sec:RVs}

Although the correlations between activity indicators are useful in terms of trying to understand some of the physics driving their long-term variations, the long-term goal is to understand, and correct for, RV variations induced by the stellar activity cycle. In Section~\ref{sub:reduction} we corrected the RV measurements for secular acceleration and known planetary signals -- and hence any remaining RV variability should primarily be driven by stellar activity (though there is the possibility that, as yet, undetected planets also contribute to the RVs). In this Section we investigate trends between the activity indicators and the RVs in the same way as we investigated the trends between the different activity indicators in Section~\ref{sec:indicators}.

\subsection{Correlations between long-term RV and activity indicator variations}

In Section~\ref{sec:indicators}, we used a seasonal binning approach to mask short-term (rotationally modulated) activity variations in order to investigate the underlying long-term changes in stellar activity. While this works well for activity indicators, for the RVs, however, we have opted for an extra degree of smoothing by using a polynomial fit. The reason for this difference is that there is still a possibility that an RV signal exists in the data that is independent from stellar activity. This additional source of RV variability may stem from, for example, unseen or poorly removed planets, or slight wavelength calibration offsets (which are seen in HARPS data from night-to-night). Therefore, these possible sources of RV variations could cause problems when performing the seasonal binning in the same way as for the activity indicators. Thus, we decided to fit the RV data (after binning any data per night) using a $4^\mathrm{th}$ order least squares polynomial fit to further smooth the data. It is then the correlation between the seasonally binned activity indicators and the smoothed RV data (sampled at the same times) that is used to investigate the relation between stellar RVs and activity indicators over long-timescales. We note that when we bin the RVs without smoothing (i.e. in the same way as done for the activity indicators), we qualitatively find the same results, albeit with a larger scatter.

 \begin{figure}
    \centering
    \includegraphics[width=\columnwidth]{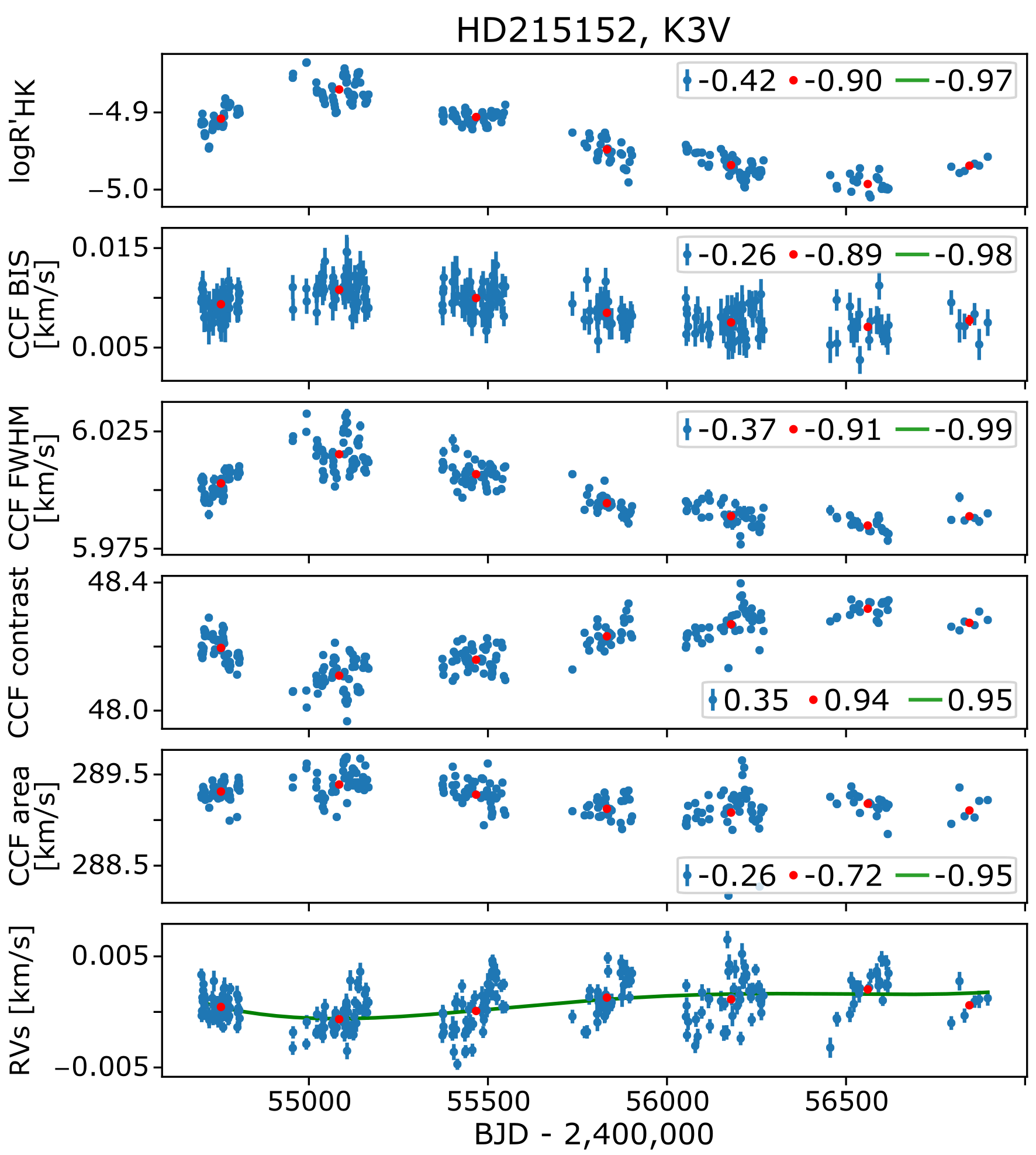}
    \caption{\label{fig:Correlation_3} Long-term activity and RV trends for the K3V star HD215152. The nightly (weighted-mean) binned data are shown in blue, the seasonally binned data are shown in red, and the green curve corresponds to the $4^\mathrm{th}$ order polynomial fit, used to smooth the RV data. The Pearson's $R$ correlation between activity indicators and RVs for each set of data are shown for each case in the legend. While the correlation of the seasonal binning and the one that used the polynomial fit are very similar, the correlation of the daily points, however, is dominated by shorter period (most likely) rotational modulation, and this weakens the long-term correlations.}
\end{figure}

Similar to Section~\ref{sec:indicators}, we use the Pearson's $R$ correlation coefficient to determine possible correlations between the RVs and the five activity indicators (log\,$R'_\mathrm{HK}$, CCF BIS, CCF FWHM, CCF contrast and CCF area). As before, we stress that our main focus is not the exact value of the correlation coefficient itself. Moreover, we verified that using the Spearman correlation coefficient does not affect our overall conclusions. Figure~\ref{fig:Correlation_3} presents an example of the long-term trends of the K3V star HD215152 both for the 5 activity indicators and the RVs. The data (binned nightly where appropriate) are shown in blue, the seasonally-binned data are in red, and the polynomial fit (to the nightly-binned data) used to smooth the RV data is displayed in green. The respective Pearson's $R$ correlation coefficients between the RVs and each of the activity indicators are shown in the inset panel for each of the different binning methods. In this case, we find that the correlation coefficients found between the activity indicators and the RVs are similar if considering the data when binned by season (as done in Section~\ref{sec:indicators}), or when we assess the correlation using the smoothed RV data. However, the correlation (or anti-correlation) between the activity indicators and the RVs become much weaker if the individual (nightly-binned) data are used. This difference is due to the rotationally modulated stellar activity signals that we are now sensitive to in the latter case, which effectively adds a source of `noise' that confuses the analysis of the longer-term activity-cycle trends. 

\begin{figure*}
    \centering
    \includegraphics[width=1\textwidth]{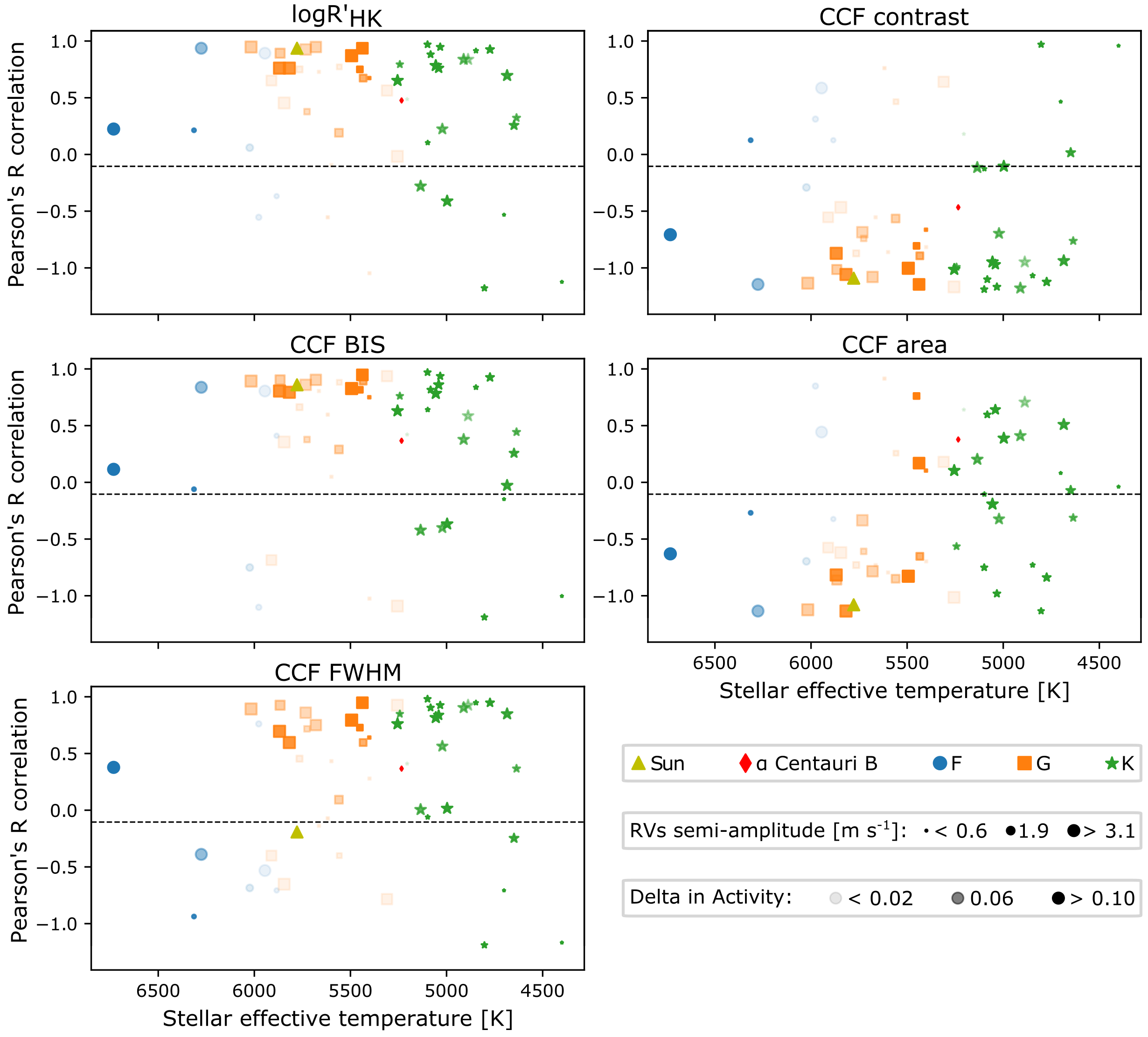}
    \caption{\label{fig:Correlation_6} Correlation between the log\,$R'_\mathrm{{HK}}$, the bisector span, the FWHM, the contrast and the area with the RVs as a function of the stellar effective temperature value for the 54 stars in our sample. The Pearson's $R$ correlation was measured between the seasonally binned activity indicators data and the smoothed seasonally binned RV data. The colour scheme and transparency effects are the same as used in Figure~\ref{fig:Correlation_2}. In addition, the size of the points was adjusted to represent the scale of the change of the long-term RV variations. According to this figure, a change in the correlation between the RVs and the different activity indicators occurs around mid-K stars.}
\end{figure*}

\begin{figure*}
    \centering
    \includegraphics[width=1\textwidth]{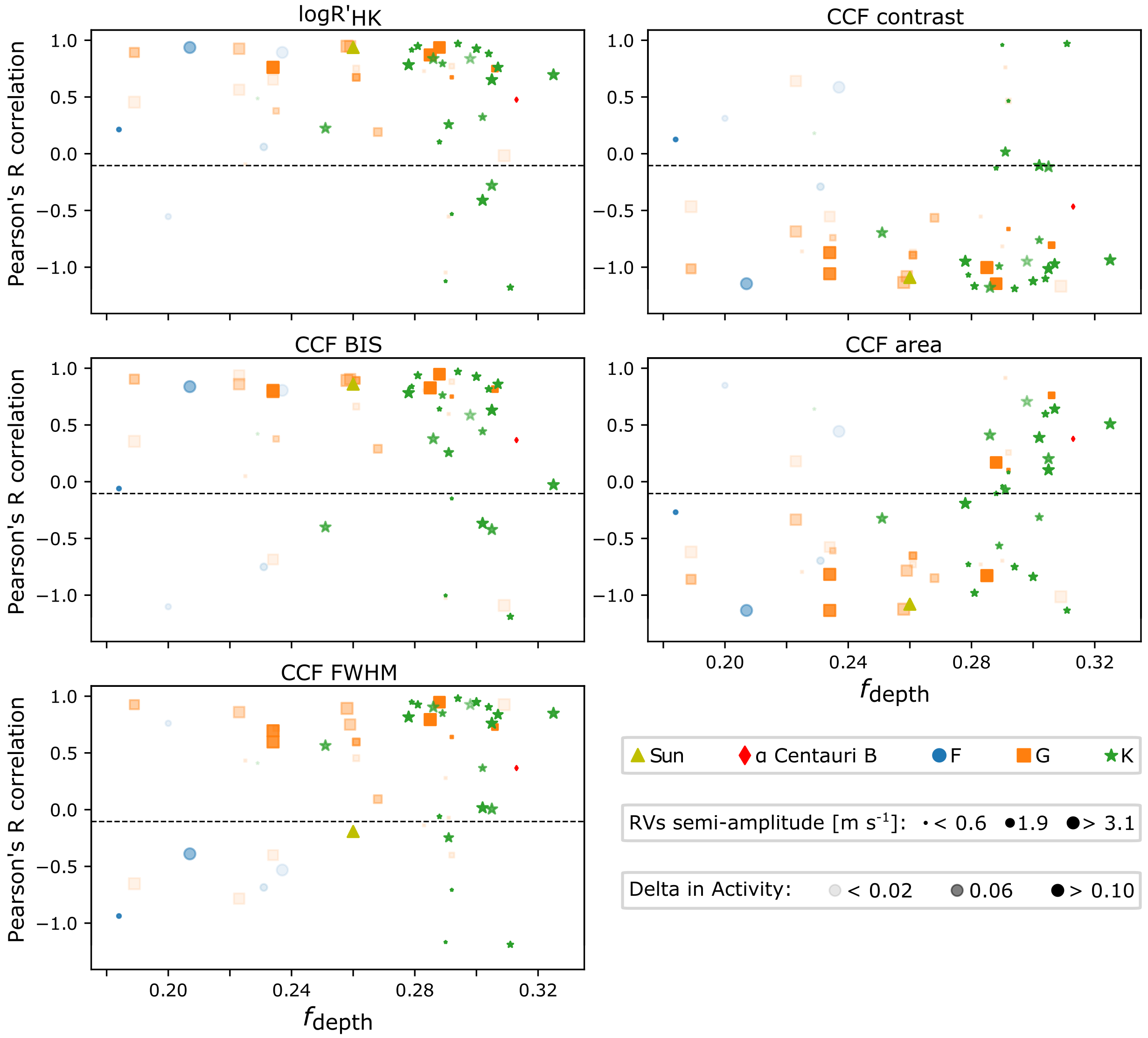}
    \caption{\label{fig:Correlation_61} Same as Figure~\ref{fig:Correlation_6}, but now as a function of fractional convective zone depth ($f_\mathrm{depth}$). Note that the x-axis is restricted to capture the bulk of the stars that exhibit the change of correlation. By using the fractional convective zone depth instead of the stellar effective temperature, the transition, observed in Figure~\ref{fig:Correlation_6} around mid-K stars, is seen here more clearly. Thus, this change of correlation seems to happen around stars with bigger convective envelopes.}
\end{figure*}

The same analysis of the long-term correlation between the seasonally binned activity indicators and the smoothed seasonally binned RVs was done for all the 54 targets and is presented in the next section. Figure~\ref{fig:Correlation_6} represents the Pearson's $R$ correlation value between activity indicators and the RVs as a function of stellar effective temperature for our 54 stars. We also present these correlations as a function of convective zone depth in Figure~\ref{fig:Correlation_61}. The Pearson’s $R$ correlation was calculated for the binned data (after smoothing the RV with a polynomial fit), and we used the same scheme for the colours and the transparency in Figures~\ref{fig:Correlation_6} and \ref{fig:Correlation_61} as in Figure~\ref{fig:Correlation_2}. In addition, we have also adjusted the size of the points to represent the change of the long-term RV variations. This was done by using a Generalised Lomb-Scargle (GLS, \citealt{zechmeister2009generalised}) periodogram, and by measuring the semi-amplitude of the best-fit sine curve of the long-term signal. Hence, a small point indicates that the RVs show a small RV change over the magnetic cycle and, therefore, we should treat these cases with care. In the following Section we investigate the changes in the behaviour of the correlations seen in Figures~\ref{fig:Correlation_6} and \ref{fig:Correlation_61}.

\subsection{Overall picture and physical interpretation}

In Section~\ref{transition} we showed that transitions and/or bifurcations occur in the long-term behaviour of the correlations between different activity indicators. These can be seen as a function of stellar effective temperature, and become even more apparent when we consider the correlations between activity indicators as a function of convective zone depth. Since these transitions are likely driven due to changes in the physical manifestations of magnetic activity on the stellar surfaces, we wish to assess whether this also has an impact on how the long-term RV variations respond to changes in the stellar activity indicators as a function of fundamental stellar parameters.

\subsubsection{The case of hotter stars: F- and early G-type stars}

For F- and early G-type stars, we observe a positive correlation between the RVs and the log\,$R'_\mathrm{{HK}}$ and bisector span, while conversely we find an anti-correlation between the RVs and the CCF contrast and CCF area. These observations are in agreement with Figure~\ref{fig:Correlation_2} and correspond to what we would expect for quiet solar-type stars. For example, as plage coverage increases with magnetic activity, magnetic elements will inhibit convective motions resulting in the suppression of convective blue-shift that results from granulation \citep{Meunier2010}. Since on long timescales the RV signal is mostly due to the inhibition of the convective blue-shift in plage \citep{SaarFischer2000,Meunier2010}, when the magnetic activity increases, so does the apparent stellar RV, leading to a strong positive correlation on longer timescales as seen in Figure~\ref{fig:Correlation_6}. 

The correlation between the RVs and FWHM for the F- and early G-type stars mirrors what we observe for the Sun. Indeed, most of the hotter stars present a small ($\Delta_{\mathrm{act}}$) and a weak (anti-)correlation. From our previous discussion of the correlation between the CCF FWHM and log\,$R'_\mathrm{HK}$ for the Sun (see Figure~\ref{fig:Correlation_1} and Section~\ref{PhysicalMeaning}), we therefore infer that these hotter stars are plage-dominated rather than spot-dominated. This is likely primarily due to their shallow convective envelopes (see Figure~\ref{fig:Correlation_61}). Thus, the FWHM, which is predominantly impacted by the presence of spots, does not contribute strongly to the RV variations, thereby explaining the propensity of early-type stars in our sample (including the Sun on approach to solar minimum) to show a weak (anti-)correlation between RVs and FWHM.

\subsubsection{The case for late G- and early K-type stars}

For late G- and early K-type stars, Figure~\ref{fig:Correlation_6} shows that the correlations between RVs and the stellar activity indicators follows a similar trend to that between the activity indicators and log\,$R'_{\mathrm{HK}}$. The log\,$R'_\mathrm{{HK}}$, BIS and contrast show the same correlation sign as seen for the hotter stars, again likely due to the suppression of convective blue-shift. The difference between the late G- and early K-type stars and the hotter stars is found for the FWHM, where the majority of stars now present a strong correlation. Assuming that the FWHM is an indicator for spots, this would indicate that the quiet late G- and early K-type stars in our sample have an increase in their spot-coverage, and that these spots have a stronger effect on the long-term RVs compared to that observed for the Sun (see Figure~\ref{fig:Correlation_1}). 

Finally, for the correlation between the CCF area and the RVs, we see the same bifurcation in Figure~\ref{fig:Correlation_6} occurring around late G- and K-type stars as also noticed in Figure~\ref{fig:Correlation_2}. While for hotter stars the CCF area is dominated by the contrast, as seen for the Sun, in the case of late G- and K-type stars the CCF area presents both anti-correlations and correlations with RVs. As explained in Section~\ref{PhysicalMeaning}, this change in correlation arises due to a difference in the activity and metallicity level among the stars. As before, when these correlations are presented as a function of convective zone depth (see Figure~\ref{fig:Correlation_61}) we see a clear transition in the correlation behaviour, rather than a bifurcation. Following from this, we therefore hypothesize that the behaviour of the correlation between the CCF area and the RVs is explained by stars of two broadly different activity levels:

\begin{enumerate}
\item lower-activity, poorer-metallicity, plage-dominated stars (with shallower convective zones) that display an anti-correlation in CCF area versus RVs as the CCF area changes are driven by the contrast, and
\item more active stars, with richer metallicity and with higher spot filling factors (and deeper convective zones), where a correlation between CCF area and RVs is driven by changes in the spot-affected FWHMs.
\end{enumerate}

\subsubsection{The case of cooler stars: mid-K stars}
\label{convectiveredshift}

As can be observed in Figure~\ref{fig:Correlation_6}, a few mid- to late-K stars exhibit a strong (anti-)correlation that is opposite to what is seen for the hotter stars in our sample. This is also the case for HD215152, previously shown in Figure~\ref{fig:Correlation_3}. Indeed, while most stars show a strong correlation between log\,$R'_\mathrm{{HK}}$ and the observed RVs, HD215152 shows a strong anti-correlation. This difference could be interpreted as evidence of convective red-shift in some later-type stars (rather than convective blueshift, e.g. \citealt{Kurster2003}). Thus, as the convection is suppressed by increasing magnetic field, we see the suppression of this convective redshift effect (yielding a net blueshift in the measured RVs) -- leading to a reversal in the RV versus log$R'_{\rm HK}$ correlations. If convective red-shift prevails, then this may suggest that most of the photospheric absorption lines used for the RV measurements of late K- (and probably M-) type stars either form in regions of convective overshoot, or that the convective redshift is due to an opacity effect (as seen the in M star models, e.g. \citealt{Beeck2013a, Beeck2013b, Norris2017}). 

Such a signature of possible convective red-shift suppression is not unprecedented. \cite{Lovis2011} show a similar tendency, with their correlation slope between the RV and the $R'_{\rm HK}$ gradually changing from positive to negative (i.e. from correlated to anti-correlated, see their Figure~17 top and Figure~18 bottom) around 4800 K (corresponding to spectral class $\sim$K3)\footnote{One can note that their correlation slope between the RV and the $R'_{\rm HK}$ seems to bend away from negative values when more active stars are included (see \citealt{Lovis2011}, Figure~18 top). This could be due to the rotational enhancement effect \citep{Dravins1990,BruningSaar90}. Indeed, this phenomenon causes both an average RV shift to the red, and the tops of bisectors to bend redwards as $v~\sin~i$ increases (at least to a possible saturation at about 6~$km~s^{-1}$). Since stars with higher $R'_{\rm HK}$ will have higher $v~\sin~i$ on average, therefore, the rotation effect could mask the effect of the convective red-shift, thus, cancelling or reversing the trend seen in the correlation between RVs and activity indicators.}. Moreover, \cite{Gray1982} showed a steady drift in the shapes of line bisectors, from C shaped and blue-shifted in G stars to nearly vertical and near zero velocity at G8-K2. \cite{Ramirez+08} see this trend continuing (the $v^{\rm plat}$ in their Table 3). Though, for a star in common with \cite{Gray+92}, $\sigma$ Dra, $v^{\rm plat}$ is shifted by -70~$m~s^{-1}$. If we apply this correction to the \cite{Ramirez+08} values, we find net {\it red-shifts} in stars cooler than $T_{\rm eff} = 5000$ K. Suppressed convection in these objects would thus yield {\it blue-shifts}, as we observe here in some mid-K dwarfs. Along these lines, \cite{SaarBruning90} show the slope of the bisector (defined as a linear fit to the bisector when the line profile is rotated -90$^o$) goes through zero and becomes negative (i.e. has a redwards tilt at top) at B - V $\approx$~0.90 in inactive (log\,$R'_\mathrm{{HK}} \leq -4.75$) stars, again corresponding to spectral class $\sim$K3. Thus, there would seem to be a gradual drift in bisector (and hence convective) properties from the G stars (blue-shifted C shapes), to near zero (near vertical for early-K), into mid-K (red-shifted, inverse C shapes). This change in convective properties may even persist into M-type stars \citep{Kurster2003}.

Finally, as we saw previously in Section~\ref{PhysicalMeaning}, analysing the correlations between activity indicators as a function of convective zone depth again yields important insights. For the K-stars in our sample, instead of a scatter or bifurcation in the correlations between RVs and activity indicators, Figure~\ref{fig:Correlation_61} displays a more gradual transition in the stellar behaviour. This is particularly evident when comparing the RVs versus CCF area in Figure~\ref{fig:Correlation_6} to those of Figure~\ref{fig:Correlation_61}. Thus, for hotter stars with shallower convective envelopes, a strong (anti-)correlation between the activity indicators and the RVs due to the convective blue-shift is observed. However, as we move to cooler stars with deeper convective envelopes, a gradual change in the correlation between activity indicators and RVs appear, possibly due to the emergence of convective red-shift. At a certain point, the convective blue- and red-shifts `cancel' one another, removing most of the correlation we see between the RVs and the activity indicators. This would provide a natural explanation for the weak correlation seen for some of the stars in Figure~\ref{fig:Correlation_61}. Finally, for even later-type stars, the convective blue-shift would then be fully replaced by the convective red-shift, explaining the strong opposite-sense of the correlations we see between the activity indicators and the RVs, as observed, for example, with HD215152 (see Figure~\ref{fig:Correlation_3}).

\subsubsection{Other effects that could impact the observed correlations}

It is pertinent to consider other effects that may potentially confuse and/or drive different correlations. For instance, phase shifts of a few days have been previously been identified between measured RVs and stellar activity indicators (e.g. \citealt{Bonfils2007, Santos2014, Cameron2019}). Such phase shifts would introduce a lag between the variations of the different measurements and hence act to weaken the measured correlation coefficients presented in this paper. However, these phase shifts tend to be between 1 to 3~days and hence our approach to binning the data to remove short term variability should mask such shifts.

Meridional flows (e.g. \citealt{Makarov2010, Meunier2020}) have been predicted to have an important impact on RVs over the course of the stellar activity cycle. By reconstructing integrated solar RVs, \cite{Meunier2020} showed that such flows induce RV signals of the order of 0.7~m~s$^{-1}$ for stars viewed edge-on (where the inclination of the stellar rotation axis is 90$^{\circ}$), and 1.6~m~s$^{-1}$ for stars viewed pole-on. Importantly, the effects from meridional flows exhibit a time lag with the magnetic cycle (about 2 years for the Sun), and this lag could substantially degrade the correlation between the RVs and log\,$R'_\mathrm{{HK}}$. Because the presence of multi-cellular pattern in F- and early G-type stars decreases the meridional flow contribution to the RV signal, we expect to see a stronger degradation of the RV versus activity indicator correlations for late G- and K-type stars. Looking at Figure~\ref{fig:Correlation_6}, we do indeed see a greater scatter in these correlations for K-type stars. While we discussed a possible physical interpretation of this in Section~\ref{convectiveredshift}, meridional flow could have an effect on the data as well, and could provide an additional explanation for what is seen. We note that a transition from a strong correlation to a strong anti-correlation (as seen in Figure~\ref{fig:Correlation_61}) would not be expected, unless the time lag for the anti-correlated stars was of the order of half the activity-cycle length. In addition, we examined the correlations for evidence of any time lag between the RVs and the activity indicators. While the gaps in the sampling makes this more complicated, no systematic and/or significant time lag between the RVs and the log\,$R'_\mathrm{{HK}}$ was found in our data.

Finally, the combination over time of geometrical effects (for example, the interplay between the inclination of the stellar rotation axis and the latitude of active region emergence) with activity level variation can also weaken the correlation between the RVs and the stellar activity indicators \citep{Meunier2019}. Furthermore, the weakening of the correlation seems to be amplified by the presence of meridional flows and its associated time lag \citep{Meunier2020}.

There are, potentially,
other surface flows that could further impact observed RV--activity indicator correlations. \cite{Haywood+2020} cite several that are related to active regions including: penumbral flows, moat flows (that are radially directed outside of the penumbra, within the surrounding plage), and ``active region inflows" \citep{Braun2019}. The latter, in particular, may be of importance as they surround active regions but seem rather asymmetric, mostly directed towards the active region from poleward directions with a strong counter-rotational component. Such a flow would affect RVs over both the rotational and activity cycle timescales, and are also expected to produce an apparent time lag with respect between activity indicators.


\section{Correcting for activity cycle RVs variation}
\label{sec:Discussion}

\subsection{Removing the long-term RV trend driven by activity}

To confirm and measure the mass of an Earth-like planet in Earth-like orbit (or other similarly long-period planets) an observation spanning at least $\sim$2 years will be necessary in order to recover and confirm a repeating signal. Removing the long-term RV trends caused by stellar activity cycles as observed in Figure~\ref{fig:Correlation_3} may prove crucial to reveal the presence of long-period planets. To achieve this, we investigated a simple method to remove these activity-cycle related RV variations. We focused on the stars for which at least one of their activity indicators showed a strong (anti-)correlation ($|R|>$~0.6) with the measured RVs. Using a $4^\mathrm{th}$ order polynomial fit for each strong (anti-)correlated activity indicator using the daily binned data, we then scaled the fit to match the amplitude of the long-term RV trend, and then subtracted the fit from the RV data using an optimal subtraction method. The optimal subtraction process selects the scale factor of the fit such that the residuals are minimised once the scaled fit is subtracted from the data. Thus, in this process we are simply using the different activity indicators as a proxy for the stellar-activity driven RV variations. Finally, we analysed these ``corrected'' RVs by comparing the GLS periodogram before and after applying the correction to the RVs. For each target, we considered the removal of the long-term RV trend using activity indicators as successful if the GLS periodogram of the ``corrected'' peak at long periods (over 3000~days) was lower than the one before the removal of the scaled activity indicator fit. Moreover, this ``corrected'' peak also needed to be lower than the 5\% false alarm probability (FAP) level of the periodogram. A 5\% FAP level was chosen as we want to ensure that any spurious signal is below a 2~$\sigma$ detection limit. This allowed us to determine whether power in the GLS periodogram at long periods (induced by stellar activity) was significantly reduced. 

Figure~\ref{fig:Correlation_5} shows an example of the results of this process for HD215152. As we are focusing on long-term signals, we have reduced the opacity of the short-period portion of the periodograms (for periods less than 600 days) to highlight the region of interest. In blue we show the periodogram of the RVs before being modified, and orange shows the RV periodogram after removing the scaled activity indicator fit. Due to their strong correlation with the RVs, the scaled fit from all the activity indicators remove the long-term trend seen in the RVs to some degree, although they do not affect the prominent $\sim$365 day period. This is to be expected, as this peak, which corresponds to the spurious one-year signal in the HARPS data found by \cite{Dumusque2015b}, is not due to activity and therefore is not expected to be removed after correcting the RVs. 

After applying this method on our 54 targets, the long-term RV trend was successfully removed for 40 targets, for at least one activity indicator. Table~\ref{tab:3} presents the results of this method. First, we show the number of stars with a strong ($|R|>$~0.6) correlation between each of the activity indicators and the RVs. We then provide the number of stars for which this activity indicator was successful in correcting the long-term RV trends. For instance, 34 targets show a strong correlation between log\,$R'_\mathrm{{HK}}$ and the RVs. For 32 of them, the log\,$R'_\mathrm{{HK}}$ can be used to correct the long-term RVs signal. While we presented an example, Figure~\ref{fig:Correlation_3}, where all activity indicators were strongly (anti-)correlated with RVs and correctly removed the long-term RV trend, we found that this is rarely the case. For instance, the log\,$R'_\mathrm{{HK}}$ failed to remove the long-term RV variations of 8 targets (20\% of the successful 40 targets), while other activity indicators succeeded. Indeed, depending on the star, some activity indicators will show a stronger correlation than others, and hence, will better match the long-term RV trend induce by stellar activity and will prove to be better at correcting the RVs. This shows that it is extremely important to carefully check different activity indicators (and not just one indicator, like log\,$R'_\mathrm{{HK}}$, or none at all, as previously done by, for example, \citealt{Vogt2010, Kane2014, Moutou2015, Delisle2018, Udry2019}) before judging if long-term RV trends are due to a planetary signal or due to the magnetic cycle. We note that, at the moment, we are not able to find a clear relationship between which activity indicator(s) are best to correct the long-term RV trends and any fundamental property (e.g. spectral-type) of the star.

\begin{table*}
    \centering
    \begin{tabular}{c|c|c|c|c|c}
    \hline
    \hline
    	Activity Indicators  &   log\,$R'_\mathrm{{HK}}$ &   BIS  &   FWHM  &   contrast  &   area  \\
    \hline
    No. [\%] of stars that display a strong  &      &     &     &    &     \\
    correlation between the RVs and each   &   34	[63\%]   &  35	[65\%]   &  31 [57\%]   & 32 [59\%]   &  21 [39\%]   \\
    activity indicator for the 54 targets    &      &     &     &    &     \\
    \hline
    No. [\%] of stars for which long-term   &      &    &      &    &     \\
    RV trends could be removed using each   &   32 [59\%]   &   33 [61\%]  &   30 [56\%]   & 29 [54\%]   &  16 [30\%]   \\
    activity indicator for the 54 targets   &      &    &      &    &     \\
    \hline
    \hline
    \end{tabular}
\caption{The number of stars showing a strong correlation or strong anti-correlation ($|R|>$~0.6) for the different activity indicators as well as the number of stars for which we are able to remove long-term RV trends using this indicator looking at the 54 targets.}
\label{tab:3}
\end{table*}

\subsection{Detecting hidden planets}

 \begin{figure*}
    \centering
    \includegraphics[width=1\textwidth]{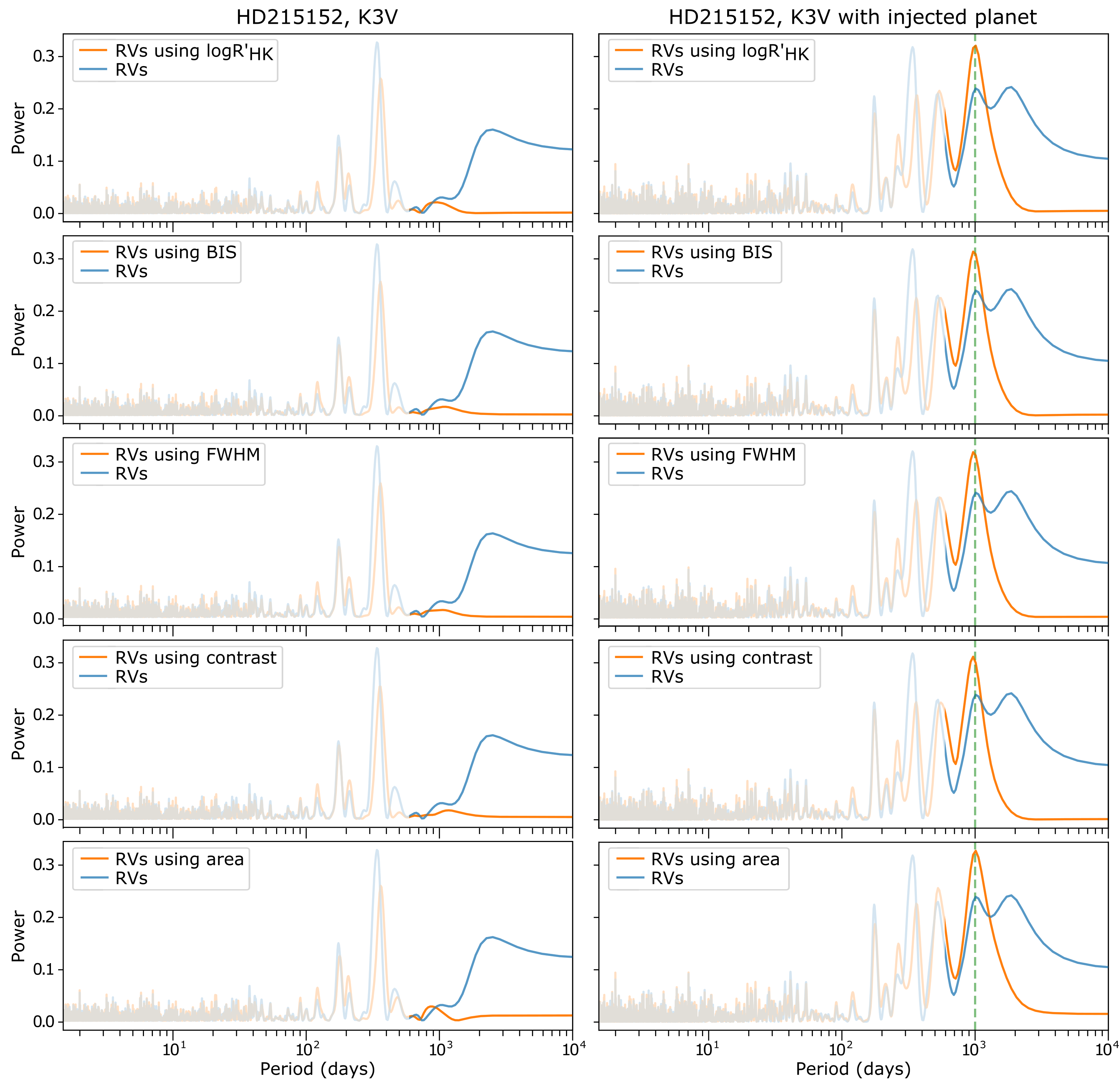}
    \caption{\label{fig:Correlation_5} {\it Left:} The GLS periodogram of the RVs of the K3V star HD215152 before (blue) and after (orange) correcting for long-term stellar activity using different activity indicators. As we want to focus on long-term trends, we have decreased the opacity for periods less than 600 days. As one can see, the long-term RV trend seen in the data has been fully removed, with the decrease of the long-term peak. {\it Right:} the GLS periodogram of the same star, but now with a fake planetary signal added into the RVs. The fake planet has a semi-amplitude of 1.3~m~s$^{-1}$ and an orbital period of 1000~days. The fake planet's period is shown with the vertically dashed green line. As expected, without removing the long-term trend, it is not possible to confidently recover the injected planet signal at the right period. However, when the long-term RV trend induced by stellar activity is removed using the best activity indicators, the planet's signal can be easily recovered at the correct period, and now produces the strongest signal in the periodogram.}
\end{figure*}

Although for many stars in the sample we are able to correct for the long-term activity effects in the RVs, we also need to investigate whether this improves our ability to detect long-term planets. In order to do so, we injected a fake circular planetary signal, of semi-amplitude 1.3~m~s$^{-1}$ with a period of 1000 days (corresponding to a Neptunian planet) in the RV data. We subsequently removed the stellar activity effects as before to look for the injected planet. We show the resulting GLS periodograms after injecting the fake planet in the data of HD215152, both before and after the correction for long-term trends, in Fig.~\ref{fig:Correlation_5}. The period at which the planet was injected is highlighted with a vertical dashed green line. As one can see, if the RVs are not corrected for the stellar activity, the planetary signal is predominantly hidden in the stellar noise (blue line), or is detected at the wrong period, compared to when the long-term trend is removed (orange line). This shows the importance of carefully removing the long-term RV trend, using the correct activity indicators, as planetary signals could be hidden inside.

We would like to caution that this simple cleaning of the long-term RVs may, however, also partially remove the planet induced RV variations if the planetary signal has the same structure as the magnetic cycle (i.e. a similar period and phase). A more detailed analysis of the signals hidden in the long-term stellar noise (for the targets presented in this paper), including new planet candidates, will be presented in a future paper using more sophisticated removal methods (Costes et al., in prep.).


\section{Conclusions}
\label{sec:conclusions}

In this paper, we analysed how the long-term activity trends for the CCF BIS, FWHM, contrast, and area correlated with log\,$R'_\mathrm{{HK}}$ and the measured RVs of 54 stars (from F-type to K-type stars) using HARPS and HARPS-N solar telescope data. In order to assess the strength of the long-term correlation between activity indicators and the log\,$R'_\mathrm{{HK}}$, seasonally binned data were analysed. In this study, we found:
\begin{itemize}
  \item The strength and sign of the correlation between both contrast and the bisector span with log\,$R'_\mathrm{{HK}}$ are similar for most spectral types.
  
  \item A transition from anti-correlation (or weak correlation, for hotter stars such as F- and early G-type stars) to a strong correlation (for cooler stars) is observed between the CCF FWHM and the log\,$R'_\mathrm{{HK}}$. We speculate that this transition can be explained by the shallow convective envelope of hot stars, resulting in a weak magnetic flux that is not strong enough to generate substantial spot-coverage. Since the FWHM mostly traces spots, the correlation between the FWHM and the log\,$R'_\mathrm{{HK}}$ is then very weak for these plage-covered stars, with a small tendency to be anti-correlated. As we move to cooler stars, spot-coverage increases, leading to a stronger correlation between the FWHM and the log\,$R'_\mathrm{{HK}}$.
  
  \item An interesting feature in the correlation of the CCF area with log\,$R'_\mathrm{{HK}}$ occurs around late G- and K-type stars. While the CCF area and log\,$R'_\mathrm{{HK}}$ are anti-correlated for hotter stars, there is a bifurcation for cooler stars, where CCF area can be correlated or anti-correlated with log\,$R'_\mathrm{{HK}}$. When considered as a function of convective zone depth (rather than stellar effective temperature) this bifurcation is transformed into a clear transition. Therefore, we can classify the cooler type stars into two groups:
  \begin{enumerate}
    \item The first group of stars (those that display a positive correlation) have deeper convective zones. These stars are also more metal rich and display higher median activity levels. We hypothesize that this leads to a higher spot-coverage resulting in an increase in the contribution of the FWHM to the overall CCF area variations, thus explaining the positive correlation between the area and the log\,$R'_\mathrm{{HK}}$.
    \item The second group of stars (those that present an anti-correlation) have shallower convective zones. These stars are also more metal poor and display lower median activity levels. We hypothesize that these are plage-dominated stars, with fewer spots. Thus, the contrast becomes the main driver of the CCF area variations, explaining the anti-correlation seen (since the contrast is anti-correlated with log\,$R'_\mathrm{{HK}}$).
  \end{enumerate}
\end{itemize}

Using our results from the correlation among activity indicators, we then studied the impact of stellar activity on the RVs of the stars. By removing any non-stellar activity induced RVs, we found strong (anti-)correlations between the long-term variations in activity indicators and the RVs for a significant portion of the stars in our sample. Our analysis shows that most of the stars present the same strong (anti-)correlations across the spectral range. This mainly results from the suppression of convective blue-shift (i.e. an increase of the magnetic activity will increase the plage coverage, thus increasing the RV signal on long timescale due to the inhibition of the convective blue-shift in plage). However, in some cases, the correlations with the RVs change as a function of the spectral type of the stars. 

\begin{itemize}
  \item The first change we notice is a transition from anti-correlation (or weak correlation) to a strong correlation between the FWHM and the RVs around F- and early G-type stars. This transition is similar to what was observed for the correlation between the FWHM and the log\,$R'_\mathrm{{HK}}$. Due to their shallow convective envelopes, the magnetic field of hotter stars is not strong enough to generate significant spots. Thus, the FWHM, which mostly traces spots, will be a less effective tracer of the long-term RV variability for hotter stars, compared to cooler stars which present a higher spot-coverage.
  
  \item A second change in the correlation is observed between the CCF area and the RVs, where a bifurcation is seen for late G- and K-type stars. Similar to the bifurcation observed between the CCF area and the log\,$R'_\mathrm{{HK}}$, this bifurcation is transformed into a clear transition when considered as a function of convective zone depth. Thus, the same two groups of stars, described earlier in the correlation between CCF area and the log\,$R'_\mathrm{{HK}}$ are also observed.
  \begin{enumerate}
    \item The first group of stars have deeper convective zones, are more metal rich stars and display higher median activity levels. We think that these stars exhibit larger spot-coverages, increasing the contribution of the FWHM to the overall CCF area, thus explaining the positive correlation between the area and the RVs.
    \item The second group of stars have shallower convective zones, are more metal poor and display lower median activity levels. These stars will mostly be plage-covered, increasing the contribution of the contrast to the overall CCF area, thus leading to an anti-correlation between the RVs and the CCF area.
  \end{enumerate}
  
  \item A final change in the correlation between activity indicators and RVs is observed for stars later than $\sim$mid-K. Instead of the correlations that one would expect due to the suppression of convective blue-shift, a transition (clearer when considered as a function of convective zone depth) in the correlation is seen for cooler stars with deeper convective zones. This ``opposite'' correlation might actually be evidence for suppression of convective red-shift. Therefore, for hotter stars with shallower convective zones, the correlation between activity indicators and RVs is mainly driven by the suppression of convective blue-shift. However, we hypothesis that the suppression of convective red-shift is progressively emerging as we move to cooler stars with deeper convective zones, cancelling the correlation between activity and the RVs. Finally, this suppression of convective red-shift becomes the dominant driver, thus leading to the ``opposite'' correlations observed.
\end{itemize}

Finally, using the strong correlations observed between some activity indicators and the RVs, an attempt to remove RVs induced by stellar activity was proposed. After identifying for each star which activity indicators were the most reliable, we were able, using a scaled activity indicator fit, to fully remove the long-term RVs induced by stellar activity for 40 targets. Removing this stellar noise will be a crucial element in the search of Earth-analogue and longer-period planet signals. It is important to note that the strength in the correlation between activity indicators and RVs can vary for each star. Thus, the activity indicators used to correct the change seen in the long-term RV trend must be chosen on a star-by-star basis. In order to explicitly demonstrate the importance of removing the long-term stellar activity noise, we injected a fake planetary signal in the data. Using the data from HD215152, we showed that a Neptunian planetary signal with a period of 1000~days could only be retrieved at the right period when the RVs induced by stellar activity were removed. This proves that removing long-term RV induced by stellar activity is necessary in order to detect exoplanets hidden in the stellar activity noise. A more detailed analysis of the signals hidden in the long-term stellar noise will be presented in a future paper (Costes et al., in prep.).

\section*{Acknowledgements}

AM acknowledges support from the senior Kavli Institute Fellowships. MNG acknowledges support form MIT's Kavli Institute as a Juan Carlos Torres Fellow. JSJ acknowledges support by FONDECYT grant 1201371, and partial support from CONICYT project Basal AFB-170002. CAW acknowledges support from Science and Technology Facilities Council (STFC) grant ST/P000312/1. SHS is grateful for support from NASA Heliophysics LWS grant NNX16AB79G. DFP acknowledges support from NASA award number NNX16AD42G and the Smithsonian Institution. ACC acknowledges support from STFC consolidated grant ST/R000824/1. XD acknowledges the strong support from the Branco-Weiss Fellowship--Society in Science. This project has received funding from the European Research Council (ERC) under the European Union’s Horizon 2020 research and innovation programme (grant agreement SCORE No 851555). The HARPS-N project has been funded by the Prodex Program of the Swiss Space Office (SSO), the Harvard University Origins of Life Initiative (HUOLI), the Scottish Universities Physics Alliance (SUPA), the University of Geneva, the Smithsonian Astrophysical Observatory (SAO), and the Italian National Astrophysical Institute (INAF), the University of St Andrews, Queen’s University Belfast, and the University of Edinburgh. We would like to acknowledge the excellent discussions and scientific input by members of the International Team, `Towards Earth-like Alien Worlds: Know thy star, know thy planet', supported by the International Space Science Institute (ISSI, Bern). This research has made use of the SIMBAD database, operated at CDS, Strasbourg, France. This research has made use of the VizieR catalogue access tool, CDS, Strasbourg, France. This work presents results from the European Space Agency (ESA) space mission Gaia. Gaia data are being processed by the Gaia Data Processing and Analysis Consortium (DPAC). Funding for the DPAC is provided by national institutions, in particular the institutions participating in the Gaia MultiLateral Agreement (MLA). Based on data products from observations made with ESO Telescopes at the La Silla Paranal Observatory under programme IDs: 095.C-0040, 075.D-0194, 076.C-0878, 076.C-0073, 084.C-0229, 092.C-0721, 090.C-0421, 081.C-0842, 074.C-0364, 60.A-9036, 60.A-9700, 082.C-0308, 072.D-0707, 183.D-0729, 081.D-0870, 085.C-0063, 079.D-0075, 087.C-0990, 073.C-0784, 094.C-0797, 093.C-0062, 078.C-0044, 085.C-0019, 081.C-0034, 072.C-0096, 086.C-0230, 079.C-0657, 075.D-0760, 083.C-1001, 085.C-0318, 192.C-0852, 093.C-0409, 078.C-0751, 082.C-0427, 089.C-0497, 091.C-0034, 082.C-0212, 183.C-0972, 191.C-0873, 082.C-0718, 089.C-0732, 094.C-0894, 095.C-0551, 079.C-0170, 191.C-0505, 190.C-0027, 075.C-0234, 079.C-0681, 074.D-0380, 089.C-0050, 086.C-0284, 088.C-0662, 072.C-0513, 072.C-0488, 074.C-0012, 090.C-0849, 185.D-0056, 082.C-0315, 081.C-0802, 077.C-0530, 080.C-0032, 076.D-0130, 078.C-0833, 088.C-0011, 087.C-0831, 183.C-0437, 087.D-0511, 081.D-0065, 092.C-0579, 074.D-0131, 078.D-0071, 091.C-0936, 073.D-0038, 080.D-0086, 080.D-0347, 077.C-0364, 075.C-0332, 188.C-0265, 075.D-0800, 180.C-0886.

\section*{Data Availability}
The datasets were derived from sources in the public domain: [ESO~archive: \url{http://archive.eso.org/wdb/wdb/adp/phase3_main/form}, SIMBAD: \url{http://simbad.u-strasbg.fr/simbad/}, VizieR: \url{https://vizier.u-strasbg.fr/viz-bin/VizieR}, GAIA archive: \url{https://gea.esac.esa.int/archive/}, exoplanet.eu: \url{http://exoplanet.eu/catalog/}, NASA exoplanet archive: \url{https://exoplanetarchive.ipac.caltech.edu/}, allesfitter: \url{https://github.com/MNGuenther/allesfitter}, JLP Horizons: \url{https://ssd.jpl.nasa.gov/horizons.cgi}, EZ-Web: \url{http://www.astro.wisc.edu/~townsend/static.php?ref=ez-web}].




\bibliographystyle{mnras}
\bibliography{paper} 





\appendix
\section{Properties of the stars used}
\begin{table*}
    \centering
    \begin{tabular}{l|c|c|c|c|c|c|c|c|c|c|c}
    \hline
    \hline
    Name	&	Spectral	&	Number of	&	Number of	&	Time range	&	log\,$R'_\mathrm{{HK}}$	&	log\,$R'_\mathrm{{HK}}$	&	log\,$R'_\mathrm{{HK}}$	&	Metallicity		&	B-V	&	Teff	&	Mass		\\	
	&	type	&	spectra	&	nights	&	[days]	&	median	&	 min	&	max	&	[Fe/H]		&		&	[K]	&	\(M_\odot\)		\\	
	\hline																									
HD103774	&	F6V	&	119	&	108	&	2760	&	-4.868	&	-5.045	&	-4.758	&	0.29	$^{3}$	&	0.490	&	6732	$^{3}$	&	1.400	$^{3}$	\\
HD89839	&	F7V	&	69	&	54	&	2671	&	-5.021	&	-5.219	&	-4.850	&	0.04	$^{4}$	&	0.530	&	6314	$^{4}$	&	1.224	$^{6}$	\\
HD22879	&	F7/8V	&	114	&	59	&	3379	&	-4.931	&	-4.945	&	-4.922	&	-0.82	$^{4}$	&	0.560	&	5884	$^{4}$	&	0.782	$^{6}$	\\
HD95456	&	F8V	&	244	&	98	&	3743	&	-5.023	&	-5.053	&	-4.965	&	0.16	$^{4}$	&	0.520	&	6276	$^{4}$	&	1.288	$^{6}$	\\
HD65907A	&	F9.5V	&	744	&	207	&	1933	&	-4.942	&	-4.956	&	-4.926	&	-0.31	$^{4}$	&	0.570	&	5945	$^{4}$	&	0.915	$^{6}$	\\
HD1581	&	F9.5V	&	3127	&	254	&	3129	&	-4.951	&	-4.964	&	-4.939	&	-0.18	$^{4}$	&	0.570	&	5977	$^{4}$	&	0.992	$^{6}$	\\
HD7449	&	F9.5V	&	121	&	95	&	3819	&	-4.875	&	-4.910	&	-4.845	&	-0.11	$^{4}$	&	0.600	&	6024	$^{4}$	&	1.042	$^{6}$	\\
HD73524	&	G0V	&	167	&	96	&	2669	&	-5.025	&	-5.057	&	-4.978	&	0.16	$^{4}$	&	0.600	&	6017	$^{4}$	&	1.150	$^{6}$	\\
HD20807	&	G1V	&	235	&	61	&	4099	&	-4.898	&	-4.926	&	-4.857	&	-0.23	$^{4}$	&	0.600	&	5866	$^{4}$	&	0.935	$^{6}$	\\
HD96700	&	G1V	&	357	&	208	&	2620	&	-4.957	&	-4.984	&	-4.932	&	-0.18	$^{4}$	&	0.610	&	5845	$^{4}$	&	0.936	$^{6}$	\\
HD10180	&	G1V	&	327	&	242	&	3804	&	-5.006	&	-5.052	&	-4.986	&	0.08	$^{4}$	&	0.630	&	5911	$^{4}$	&	1.078	$^{6}$	\\
HD45184	&	G2V	&	306	&	166	&	3825	&	-4.911	&	-4.989	&	-4.858	&	0.04	$^{4}$	&	0.620	&	5869	$^{4}$	&	1.048	$^{6}$	\\
HD38858	&	G2V	&	211	&	91	&	4114	&	-4.912	&	-4.934	&	-4.880	&	-0.22	$^{4}$	&	0.640	&	5733	$^{4}$	&	0.898	$^{6}$	\\
HD189567	&	G2V	&	647	&	228	&	3602	&	-4.914	&	-4.947	&	-4.868	&	-0.24	$^{4}$	&	0.640	&	5726	$^{4}$	&	0.875	$^{6}$	\\
HD146233	&	G2V	&	5249	&	106	&	3052	&	-4.930	&	-4.988	&	-4.878	&	0.04	$^{4}$	&	0.650	&	5818	$^{4}$	&	1.031	$^{6}$	\\
Sun	&	 G2V	&	42510	&	581	&	707	&	-4.988	&	-5.006	&	-4.958	&	0.00		&	0.656	&	5778		&	1.000		\\
HD136352	&	G3V	&	674	&	220	&	2668	&	-4.952	&	-4.974	&	-4.941	&	-0.34	$^{4}$	&	0.650	&	5664	$^{4}$	&	0.842	$^{6}$	\\
HD1461	&	G3V	&	465	&	225	&	3818	&	-5.026	&	-5.059	&	-5.000	&	0.19	$^{4}$	&	0.680	&	5765	$^{4}$	&	1.065	$^{6}$	\\
HD106116	&	G5V	&	120	&	111	&	3746	&	-5.011	&	-5.039	&	-4.969	&	0.14	$^{4}$	&	0.650	&	5680	$^{4}$	&	1.012	$^{6}$	\\
HD90156	&	G5V	&	145	&	109	&	2623	&	-4.957	&	-4.970	&	-4.947	&	-0.24	$^{4}$	&	0.680	&	5599	$^{4}$	&	0.855	$^{6}$	\\
HD20794	&	G6V	&	6185	&	512	&	4118	&	-4.986	&	-5.002	&	-4.967	&	-0.40	$^{4}$	&	0.710	&	5401	$^{4}$	&	0.786	$^{6}$	\\
HD115617	&	G6.5V	&	1186	&	200	&	3362	&	-4.993	&	-5.015	&	-4.959	&	-0.02	$^{4}$	&	0.700	&	5558	$^{4}$	&	0.918	$^{6}$	\\
HD59468	&	G6.5V	&	527	&	150	&	4198	&	-5.003	&	-5.034	&	-4.988	&	0.03	$^{4}$	&	0.710	&	5618	$^{4}$	&	0.954	$^{6}$	\\
HD161098	&	G8V	&	129	&	111	&	2322	&	-4.904	&	-4.941	&	-4.858	&	-0.27	$^{4}$	&	0.670	&	5560	$^{4}$	&	0.837	$^{6}$	\\
HD10700	&	G8V	&	10393	&	419	&	3729	&	-4.959	&	-4.971	&	-4.948	&	-0.52	$^{4}$	&	0.720	&	5310	$^{4}$	&	0.742	$^{6}$	\\
HD45364	&	G8V	&	103	&	83	&	4082	&	-4.978	&	-5.024	&	-4.920	&	-0.17	$^{4}$	&	0.760	&	5434	$^{4}$	&	0.845	$^{6}$	\\
HD20003	&	G8V	&	183	&	165	&	3379	&	-5.001	&	-5.054	&	-4.864	&	0.04	$^{4}$	&	0.770	&	5494	$^{4}$	&	0.928	$^{6}$	\\
HD69830	&	G8V	&	705	&	231	&	4205	&	-5.002	&	-5.030	&	-4.912	&	-0.06	$^{4}$	&	0.790	&	5402	$^{4}$	&	0.873	$^{6}$	\\
HD157172	&	G8.5V	&	126	&	115	&	3079	&	-4.992	&	-5.042	&	-4.898	&	0.11	$^{4}$	&	0.780	&	5451	$^{4}$	&	0.933	$^{6}$	\\
HD71835	&	G9V	&	109	&	99	&	3063	&	-4.924	&	-5.032	&	-4.837	&	-0.04	$^{4}$	&	0.780	&	5438	$^{4}$	&	0.887	$^{6}$	\\
HD20781	&	G9.5V	&	226	&	198	&	3384	&	-5.038	&	-5.091	&	-4.999	&	-0.11	$^{4}$	&	0.820	&	5256	$^{4}$	&	0.832	$^{6}$	\\
HD39194	&	K0V	&	268	&	239	&	3248	&	-4.951	&	-4.989	&	-4.914	&	-0.61	$^{4}$	&	0.770	&	5205	$^{4}$	&	0.709	$^{6}$	\\
HD26965	&	K0V	&	543	&	63	&	4100	&	-4.965	&	-5.006	&	-4.864	&	-0.36	$^{4}$	&	0.820	&	5098	$^{4}$	&	0.750	$^{6}$	\\
HD72673	&	K1V	&	451	&	101	&	3456	&	-4.903	&	-4.971	&	-4.865	&	-0.41	$^{4}$	&	0.790	&	5243	$^{4}$	&	0.751	$^{6}$	\\
HD13060	&	K1V	&	90	&	82	&	3810	&	-4.842	&	-4.986	&	-4.727	&	0.02	$^{4}$	&	0.790	&	5255	$^{4}$	&	0.865	$^{6}$	\\
HD109200	&	K1V	&	740	&	355	&	3365	&	-4.952	&	-5.009	&	-4.883	&	-0.35	$^{4}$	&	0.850	&	5056	$^{4}$	&	0.744	$^{6}$	\\
AlfCenB	&	K1V	&	18698	&	309	&	2697	&	-4.932	&	-5.030	&	-4.844	&	0.16	$^{2}$	&	0.880	&	5234	$^{2}$	&	0.870	$^{2}$	\\
HD85390	&	K1.5V	&	96	&	86	&	4154	&	-4.951	&	-5.011	&	-4.881	&	-0.09	$^{4}$	&	0.850	&	5135	$^{4}$	&	0.816	$^{6}$	\\
HD13808	&	K2V	&	246	&	203	&	3368	&	-4.898	&	-4.983	&	-4.707	&	-0.21	$^{4}$	&	0.850	&	5033	$^{4}$	&	0.771	$^{6}$	\\
HD144628	&	K2V	&	260	&	155	&	3659	&	-4.933	&	-4.988	&	-4.891	&	-0.45	$^{4}$	&	0.850	&	5022	$^{4}$	&	0.715	$^{6}$	\\
HD192310	&	K2V	&	1579	&	320	&	1917	&	-5.011	&	-5.046	&	-4.856	&	-0.03	$^{4}$	&	0.910	&	5099	$^{4}$	&	0.822	$^{6}$	\\
HD82516	&	K2V	&	88	&	81	&	4158	&	-4.957	&	-5.007	&	-4.825	&	0.02	$^{4}$	&	0.910	&	5041	$^{4}$	&	0.826	$^{6}$	\\
HD101930	&	K2V	&	67	&	59	&	4108	&	-5.000	&	-5.050	&	-4.920	&	0.16	$^{4}$	&	0.910	&	5083	$^{4}$	&	0.868	$^{6}$	\\
HD204941	&	K2V	&	68	&	62	&	2721	&	-4.977	&	-5.025	&	-4.890	&	-0.20	$^{4}$	&	0.910	&	4997	$^{4}$	&	0.767	$^{6}$	\\
HD104067	&	K2V	&	89	&	83	&	2271	&	-4.750	&	-4.801	&	-4.695	&	-0.04	$^{4}$	&	0.980	&	4888	$^{4}$	&	0.780	$^{6}$	\\
HD136713	&	K2V	&	54	&	49	&	3313	&	-4.811	&	-4.883	&	-4.770	&	0.08	$^{4}$	&	0.990	&	4911	$^{4}$	&	0.813	$^{6}$	\\
HD154577	&	K2.5V	&	562	&	297	&	3575	&	-4.870	&	-4.944	&	-4.783	&	-0.73	$^{4}$	&	0.890	&	4847	$^{4}$	&	0.642	$^{6}$	\\
HD40307	&	K2.5V	&	449	&	218	&	4138	&	-4.983	&	-5.023	&	-4.839	&	-0.36	$^{4}$	&	0.950	&	4774	$^{4}$	&	0.699	$^{6}$	\\
HD215152	&	K3V	&	287	&	262	&	2196	&	-4.905	&	-5.000	&	-4.825	&	-0.08	$^{4}$	&	0.990	&	4803	$^{4}$	&	0.764	$^{6}$	\\
HD65277	&	K4V	&	50	&	44	&	3026	&	-4.994	&	-5.055	&	-4.902	&	-0.30	$^{4}$	&	1.030	&	4701	$^{4}$	&	0.692	$^{6}$	\\
HD125595	&	K4V	&	137	&	108	&	1618	&	-4.789	&	-4.877	&	-4.688	&	0.09	$^{4}$	&	1.100	&	4636	$^{4}$	&	0.761	$^{6}$	\\
HD209100	&	K5V	&	4232	&	95	&	4022	&	-4.762	&	-4.828	&	-4.687	&	-0.13	$^{4}$	&	1.060	&	4649	$^{4}$	&	0.715	$^{6}$	\\
HD85512	&	K6V	&	1046	&	501	&	4158	&	-4.976	&	-5.074	&	-4.841	&	-0.26	$^{4}$	&	1.180	&	4400	$^{4}$	&	0.658	$^{6}$	\\
HD113538	&	K9V	&	89	&	76	&	2979	&	-4.853	&	-4.969	&	-4.765	&	-0.17	$^{5}$	&	1.380	&	4685	$^{5}$	&	0.740	$^{5}$	\\
    \hline
    \hline
    \end{tabular}
    Stellar effective temperature, metallicity and stellar mass references: $^{1}$\cite{Silva2012}, $^{2}$\cite{Santos2013}, $^{3}$\cite{Delgado2015}, $^{4}$\cite{Delgado2017}, $^{5}$\cite{Santos2017}, $^{6}$\cite{Delgado2019}, $^{7}$\cite{Neves2013}.
    \caption{Table of the stars used in the study of the long-term variability. The spectral type and B - V values were taken from SIMBAD data. The log\,$R'_\mathrm{{HK}}$ was measured using the calculations described in Section~\ref{sub:logRHK}.}
    \label{tab:1}
\end{table*}

\section{Results from using the `incorrect' line-mask}
 \begin{figure*}
    \centering
    \includegraphics[width=1\textwidth]{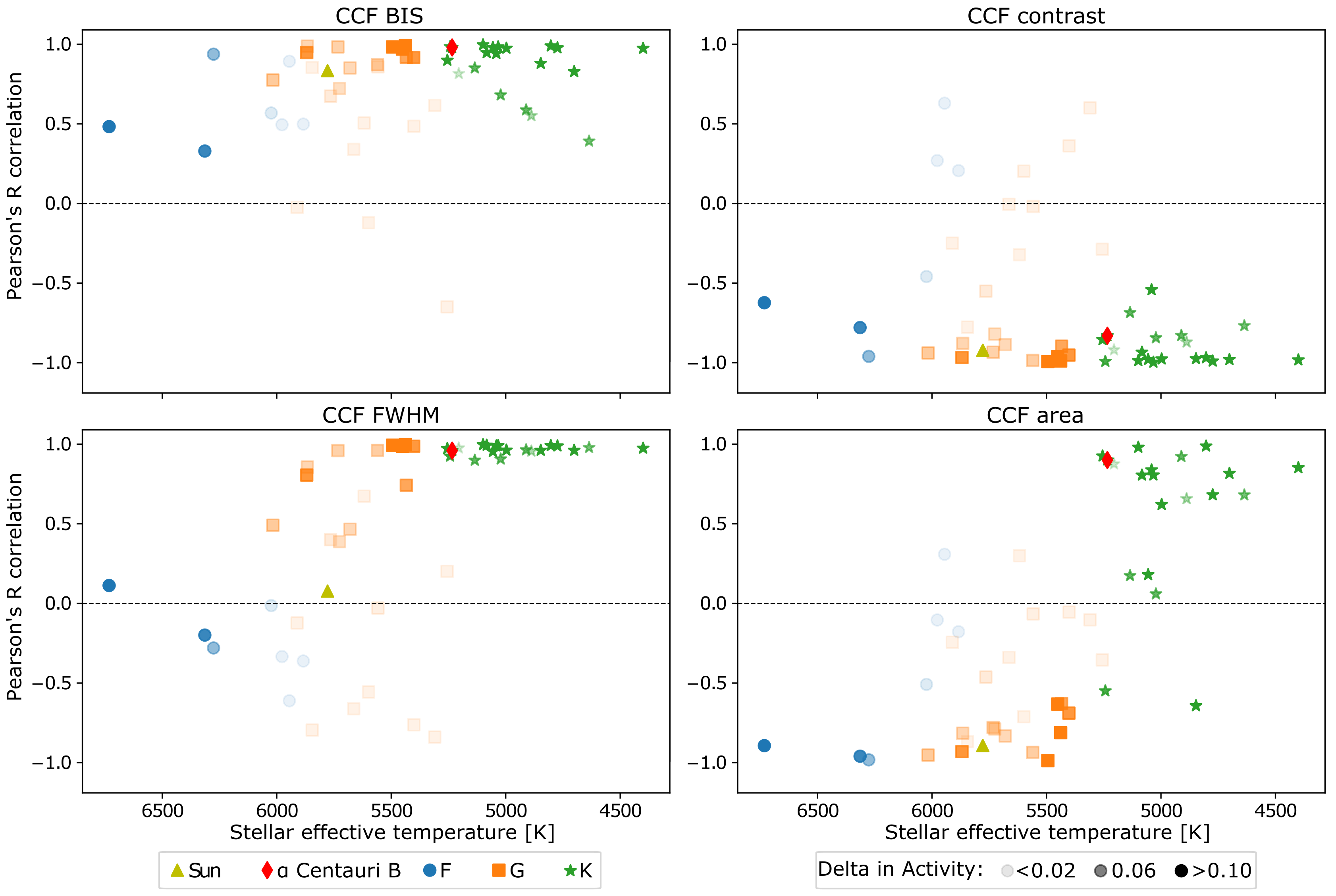}
    \caption{\label{fig:Correlation_2WM} Similar to Figure~\ref{fig:Correlation_2} but with the `incorrect' line-masks used (i.e. K5 mask used for G-type stars and G2 mask used for K-type stars). In order to simplify the comparison with Figure~\ref{fig:Correlation_2}, both the Sun and the F-type stars are presented in this plot, but were not reanalysed with a different mask.}
\end{figure*}

 \begin{figure*}
    \centering
    \includegraphics[width=1\textwidth]{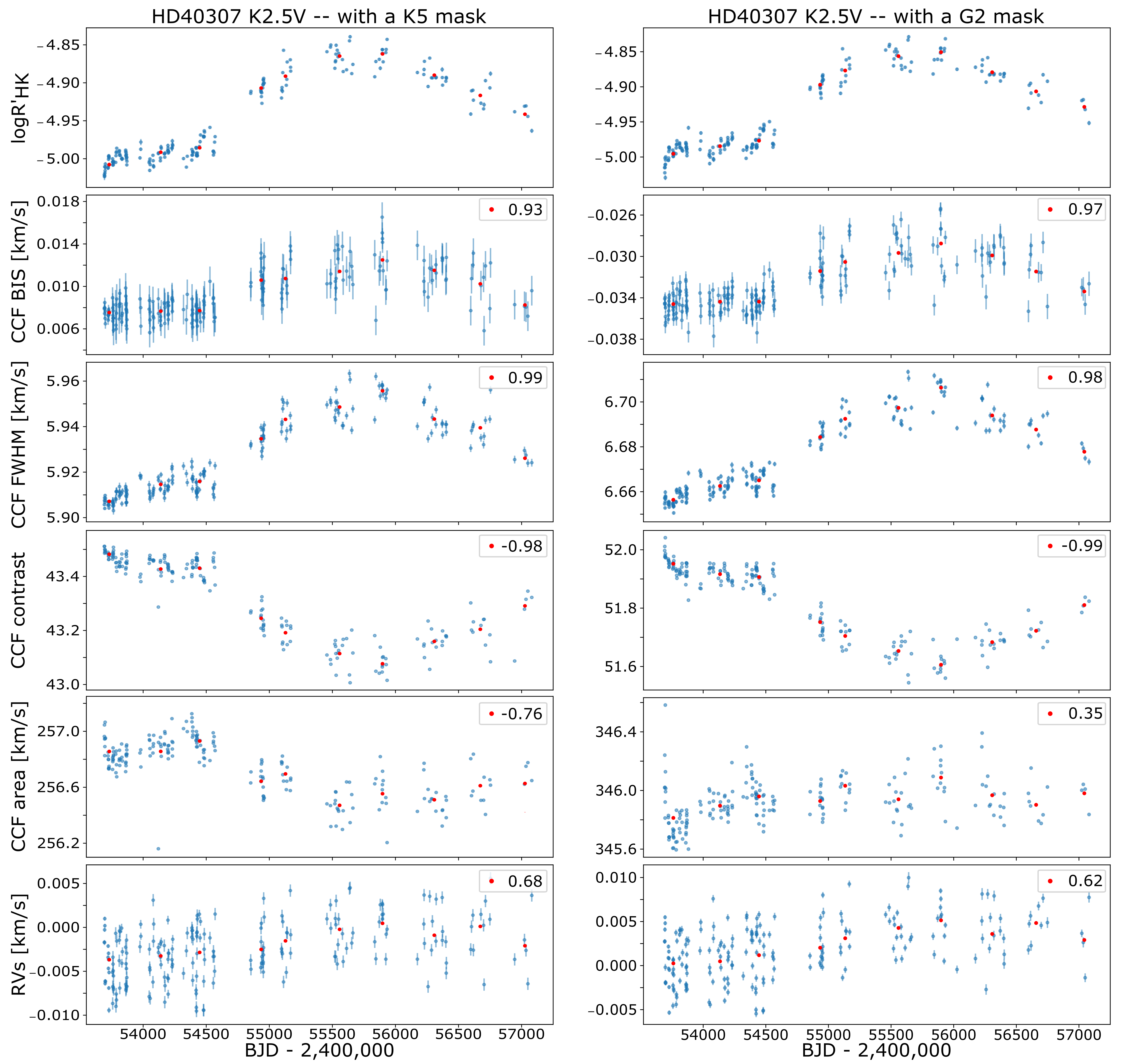}
    \caption{\label{fig:Correlation_2GK} Comparison of the long-term activity and RVs trends for the K2.5 star: HD40307, when analysed with a K5 mask (left) and with a G2 mask (right).}
\end{figure*}


\bsp	
\label{lastpage}
\end{document}